\documentclass[12pt,preprint]{aastex}
\newcommand{\lambdabar}{\lambda\kern-.9ex\raise.4ex\hbox{-}}
\newcommand{\kB}{k_{\rm B}}
\newcommand{\dxdycz}[3]{\left( \frac{ \partial #1 }{ \partial #2 }\right)_{#3}}
\newcommand{\bvec}[1]{{\mbox{\boldmath{$#1$}}}}
\newcommand{\bea}[1]{\begin{eqnarray}\label{#1}}
\newcommand{\eea}{\end{eqnarray}}
\newcommand{\eg}  {\it e.g.}        
\newcommand{\cf}  {\it cf.}         
\newcommand{\ie}  {\it i.e.}        
\newcommand{\etc} {\it etc.}
\shorttitle{Comparison of MHD and OPAL}
\shortauthors{Trampedach, D\"appen, \& Baturin}

\begin{document}
	\title{A synoptic comparison of the MHD and the OPAL\\
           equations of state}
	\author{R.\,Trampedach\altaffilmark{1,2}}
	\affil{Research School of Astronomy and Astrophysics,
		   Mt. Stromlo Observatory, Cotter Rd., Weston, ACT 2611, Australia}
	\email{art@mso.anu.edu.au}
	\author{W.\,D\"appen\altaffilmark{2}}
	\affil{Department of Physics and Astronomy, USC, Los Angeles,
           CA 90089-1342, USA}
	\email{dappen@usc.edu}
	\and
	\author{V.\,A.\,Baturin}
	\affil{Sternberg Astronomical Institute, Universitetsky 
                     Prospect 13, Moscow 119899, Russia}
	\email{vab@sai.msu.su}

	\altaffiltext{1}{Also Department of Physics and Astronomy.
		             Michigan State University, East Lansing, MI 48824, USA}
	\altaffiltext{2}{Visiting astrophycisist, Dept.~of Physics and Astronomy,
	                 University of Aarhus, DK-8000 Aarhus C, Denmark}
	
\begin{abstract}
A detailed comparison is carried out between 
two popular equations of state (EOS), the
Mihalas-Hummer-D\"appen (MHD) and the OPAL equations of state,
which have found widespread
use in solar and stellar modeling during the past two
decades.
They are
parts of two independent efforts to recalculate
stellar opacities; the international Opacity Project (OP) and the
Livermore-based OPAL project.
We examine the difference between the two equations of state
in a broad sense, over the whole applicable
$\varrho-T$~range, and for three
different chemical mixtures. Such a global comparison
highlights both their
differences and their similarities.

We find that omitting a questionable hard-sphere correction, $\tau$, to the
Coulomb interaction in the MHD formulation, greatly improves the agreement
between the
MHD and OPAL EOS. We also find signs of differences that could stem from
quantum effects not yet included in the MHD EOS, and differences
in the ionization zones that are probably caused by differences in the
mechanisms for pressure ionization.
Our analysis do not only give a clearer perception of the 
limitations of each equation of state
for astrophysical applications, but also serve as guidance for
future work on the physical issues behind
the differences. The outcome should be an 
improvement of both equations of state.
\end{abstract}
\section{Introduction}

Stellar modeling, and in particular helio- and asteroseismology, require
an equation of state and corresponding thermodynamic quantities that are
smooth, consistent, valid over a large range of temperatures and
densities, and that incorporate the most important chemical elements of
astrophysical relevance \citep[for a review see][]{jcd-wd:osc-eos}.

In astrophysics, the equation of state plays two basic roles. On the one
hand, it supplies the thermodynamic properties necessary for describing gaseous objects
such as stars and gas-planets. On the other hand it also provides the
ionization equilibria and level populations,
which can be used as the foundation for opacity calculations.
Thanks to helioseismology, the Sun
has broadened this perspective. The remarkable precision by which we have now
peered into the Sun, puts strong demands on any physics going into a solar
model. This, to such a degree, that we can
turn around the argument and use the
Sun as an astrophysical laboratory to study
Coulomb systems under conditions not yet achieved on Earth.

Although the solar plasma is only moderately non-ideal, the tight
observational constraints prompts the use of methods normally reserved for
studies of more strongly-coupled plasmas. In this way the solar experiment
addresses a much broader range of plasmas, {\eg},
Jovian planets, brown
dwarfs and low-mass stars, as well as white dwarfs \citep{cau98}.

The two equation-of-state efforts we compare in this paper are associated
with the
two leading opacity calculations of the eighties and nineties. The MHD
EOS \citep{mhd1,mhd2,mhd3} was developed for the international {\em
Opacity Project} (OP) described and summarized in the two volumes by
\citet{OP1} and \citet{OP2}. The OPAL EOS is the equation of state
underlying the OPAL opacity project at Livermore \citep[and references
therein]{fr:PLPF,rogers:OPAL-EOS}.

Another highly successful EOS address the extreme conditions in low-mass stars
and giant planets and include the transition to the fluid phase
\citep{saumon:EOS-lowM-stars,saumon:dense-EOS}.
In a trade-off between accuracy and range of validity, this EOS has so far
only been computed for H/He-mixtures, rendering it less suitable for
verification by helioseismic inversions.
The equations of state by, {\eg}, \citet{stolzmann:ThermDyn1,stolzmann:ThermDyn2}
and \citet{bi:CoulEOS} employ analytical fits to a great range of non-ideal
effects resulting in accurate, flexible and fast equation of state calculations.
The drawback of both attempts are their assumptions of complete ionization,
again making it hard to verify them by helioseismic inversions.
Comparisons with these and other EOS should, however, be
an essential part of efforts to further develop precise stellar EOS.

The OP and OPAL projects are based on two rather
different philosophies; the {\em chemical picture} and the {\em
physical picture}, respectively, as detailed in Sect.\ \ref{chemEOS} and
\ref{physEOS}.
The effect of Coulomb interactions is reviewed in Sect.\ \ref{Coulomb-corr},
and a correction, $\tau$, to these, that
seems to account for a substantial part of the differences
between the two formalisms, is explored in Sect.\ \ref{tau}.

Detailed comparisons between the MHD and OPAL EOS have proved very
useful for discovering the importance and consequences of several
physical effects \citep{eoscmp1,wd-caracas,da96}.
In Sect.\ \ref{EOSscape}, we extend these
comparisons to a systematic search in the entire $T$--$\rho$ plane, and in
Sect.\ \ref{EOSsun} we take a closer look at the EOS under solar circumstances.

The consensus of the last few years has been that
in helioseismic comparisons the OPAL
EOS is closer to the Sun than the MHD EOS \citep{GONG-Sci:sol-mod} although
both are remarkably better than earlier theories.
However, recent helioseismic inversions for
the adiabatic exponent $\gamma_1 = (\partial\ln p / \partial\ln\varrho)_{ad}$
\citep{basu:H-part-func,dimauro:high-l-suncmp}
indicates that the MHD EOS fares better than OPAL in the
upper 3\% of the sun including the ionization zones of
hydrogen and helium.

Recently the previously converging Solar abundances \citep{GN92,GN96},
have been upset by new abundance analysis \citep{AGS05}
performed on 3D simulations of
convection in the Solar surface layers. This approach avoids the free-parameters
necessary in conventional abundance analysis employing 1D atmosphere models.
The result is a lower Solar heavy element abundance causing severe disagreements with
helioseismic observations
\citep[See, e.g.,][and references therein]{bahcall:seism-AGS05}.
The consequences for equation of state issues and helioseismic measurements
are further discussed in Sect.~\ref{AGS05review}.

These new
developments again highlights the importance of competing
equation-of-state efforts and systematic comparisons such as the present.

\section{Beyond Ideal Plasmas}

The simplest model of a plasma
is a non-ionizing mixture of nuclei and electrons,
obeying the classical {\em perfect gas} law. However, an {\em ideal gas}
can be more general than a perfect gas.
Ideal only refers to the interactions between particles in the gas.
The interactions in any gas redistribute energy and momentum
between the particles, giving rise to statistical equilibrium. 
In an ideal gas these interactions do not contribute to the energy of the
gas, implying that they are point interactions.
Since the Coulomb potential is long-range in nature, and not a
$\delta$-function, real plasmas cannot be ideal.

Deviations from the perfect gas law, such as ionization,
internal degrees of freedom ({\ie}, excited states),
radiation and Fermi-Dirac statistics of electrons are all in the ideal regime.
And the
particles forming the gas can be classical or quantum, material or photonic;
as long as their interactions have infinitessimal range, the gas is still ideal.
All such ideal
effects can be calculated as exactly as desired.

The ideal picture, is however, not adequate even for the solar case.
At the solar center, an ideal-gas calculation leaves about 25\% of the gas
un-ionized. On the
other hand, the mere size of the neutral (unperturbed) atoms, do not permit
more than 7\% of the hydrogen to be unionized at these densities, provided
the atoms stay in the ground state and are closely packed. At the temperature
at the center of the Sun neither of these assumptions can possibly hold
and the mere introduction of {\em size} and {\em packed} immediately imply
interactions between the constituents of the plasma and it is therefore no
longer ideal.

In a plasma of charges, $Z$, with average inter-particle distance
$\langle r\rangle$, we define the coupling parameter, $\Gamma$, as the
ratio of average potential binding energy over mean kinetic energy $\kB T$
\begin{equation}\label{Gamma}
	\Gamma = \frac{Z^2e^2}{\kB T\langle r\rangle}\ .
\end{equation}
Plasmas with $\Gamma\gg 1$ are {\em strongly} coupled,
{\eg} the interior of white dwarfs, where coupling can become so strong
as to force crystallization.
Those with $\Gamma \ll 1$ are {\em weakly} coupled, as in 
stars more massive than slightly sub-solar.

As one can suspect, $\Gamma$ is the dimensionless coupling parameter according
to which one can classify theories. Weakly-coupled plasmas lend to systematic
perturbative ideas ({\eg} in powers of $\Gamma$), strongly coupled plasma
need more creative treatments. Improvements in the equation of state beyond
the model of a mixture of ideal gases are difficult,
both for conceptual and technical reasons.
The new treatise on stellar structure and evolution by \citet{cox-guili2}
contains a comprehensive presentation of the current state of the equation
of state.

\subsection{Chemical and Physical Picture}

The present comparison is not merely between two EOS-projects,
but also between two fundamentally different approaches to the problem.
The {\em chemical picture} is named for its foundation in the notion of a
chemical equilibrium between a set of pre-defined molecules, atoms and ions.

In the {\em physical picture} only the ``elementary'' particles of the
problem are assumed from the outset --- that is, nuclei and electrons.
Composite particles appear from the formulation.

\subsubsection{Chemical Picture: MHD EOS}
\label{chemEOS}

Most realistic equations of state that have appeared in the last 30 years
belong to the chemical picture and are based on the free-energy minimization
method. This method uses approximate statistical mechanical models (for
example the non-relativistic electron gas, Debye-H\"uckel theory for ionic
species, hard-core atoms to simulate pressure ionization via configurational
terms, quantum mechanical models of atoms in perturbed fields, {\etc}). From
these models a macroscopic free energy is constructed as a function of
temperature $T$, volume $V$, and the particle numbers $N_1, \ldots, N_m$ of the
$m$ molecules, atoms and ions (and delectrons) included in the plasma model.
At given $T$ and $V$,
this free energy is minimized subject to the stoichiometric constraints
connecting the various particle species through ionizations and dissociations.
The solution of this minimum problem then gives
both the equilibrium concentrations and, if inserted in the free energy and
its derivatives, the equation of state and the thermodynamic quantities.

Obviously, this procedure automatically guarantees thermodynamic consistency,
through the fulfillment of the Maxwell relations.
As an example, when the Coulomb pressure correction
(see Sect.\ \ref{Coulomb-corr}) to the ideal-gas contribution
originates from the free energy (and not merely as a correction to the
pressure), there will be corresponding terms in all the other thermodynamic
variables, as well as changes to the equilibrium concentrations.
This is not properly appreciated in some of the equations of state used for
modern stellar atmosphere models (all in the chemical picture), and the values
of thermodynamic derivatives will therefore depend on how they are evaluated.
This affects the value of the adiabatic temperature gradient, $\nabla_{\rm ad}$,
and hence the boundaries of convection zones.
In the physical picture, outlined below, thermodynamic consistency is ensured
in a similar way; by modeling a thermodynamic potential and evaluating all
thermodynamic quantities and derivatives from the Maxwell relations.
One major advantage of using the chemical
picture lies in the possibility to model complicated plasmas, and to obtain
numerically smooth thermodynamical quantities.

In the chemical picture, perturbed atoms must be introduced on a
more-or-less {\em ad-hoc} basis to avoid the familiar divergence of internal
partition functions \citep[see {\eg~}][]{ebeling:theory-b-states}. In
other words, the approximation of unperturbed atoms precludes the application
of standard statistical mechanics, {\ie} the attribution of a
Boltzmann-factor to each atomic state. The conventional remedy
is to modify the atomic states, {\eg} by
cutting off the highly excited states in function of density and temperature.

The MHD equation-of-state is based on an occupation probability formalism
\citep{mhd1}, where the internal partition functions $Z_{s}^{\rm int}$ of
species $s$ are weighted sums
\begin{equation}\label{Zs}
	Z_{s}^{\rm int} = \sum_i w_{is} g_{is} 
		\exp \left( - \frac{E_{is} }{\kB T} \right) \ .
\end{equation}
Here, $is$ label state $i$ of species $s$, and $E_{is}$ is the energy and
$g_{is}$ the statistical weight of that state.
The coefficients $w_{is}$ are
the occupation probabilities that take into
account charged and neutral surrounding particles. In physical terms, $w_{is}$
gives the fraction of all particles of species $s$ that can exist in state $i$
with an electron bound to the atom or ion, and $1 - w_{is}$ gives the fraction
of those that are so heavily perturbed by nearby neighbors that their states
are effectively destroyed. Perturbations by neutral particles are based on an
excluded-volume treatment and perturbations by charges are calculated from a
fit to a quantum-mechanical Stark-ionization theory
\citep[for details see][]{mhd1}.

The Opacity Project and, with it, the MHD equation-of-state
restricts itself to the case of stellar envelopes, where density is
sufficiently low that the concept of atoms makes sense.
This was the main justification for realizing the Opacity-Project in the
chemical picture and basing it on the Mihalas, Hummer, D\"appen equation of
state \citep[hereinafter MHD]{mhd1,mhd2,mhd3}.
The
Opacity Project is mainly an effort to compute accurate 
atomic data, and to use these in opacity calculations. Plasma
effects on occupation numbers are of secondary interest.

\subsubsection{Physical Picture: OPAL EOS}
\label{physEOS}

The chemical pictures heuristic separation of the atomic-physics from the
statistical mechanics is avoided in the physical picture. 
It starts out from the grand canonical
ensemble of a system of electrons and nuclei interacting through the
Coulomb potential \citep{fr:EOS2,fr:PLPF,fr:EOS}.
Bound clusters of nuclei and electrons, corresponding to ions, atoms
and molecules are
sampled in this ensemble. Any effects of the plasma environment on the
internal states are obtained directly from the statistical mechanical
analysis, rather than by assertion as in the chemical picture.

There is an impressive body of literature on the physical picture. Important
sources of information with many references are the books by
\citet{ebeling:theory-b-states}, \citet{kraeft:quant-stat-particles}, and
\citet{ebeling:therm-dyn-plasmas}. However, the majority of work on the
physical picture was not dedicated to the problem of obtaining a
high-precision equation of state for stellar interiors. Such an attempt was
made for the first time by the OPAL-team at
Lawrence Livermore National Laboratory
\citep[and references therein]{fr:PLPF,iglesias:OPAL-OP-diff,rogers:OPAL-EOS},
and used as a foundation for the OPAL opacities
\citep{iglesias-rogers:Z-opac,igl-rog:spin-orbit,iglesias-fr:Ceph-opac,iglesias:OPAL2,rog:OPAL}.

The OPAL approach avoids the {\em ad-hoc} cutoff procedures necessary in
free energy minimization schemes. The method also provides a systematic
procedure for including plasma effects in the photon absorption
coefficients.
An effective potential
method is used to generate atomic data which have an accuracy similar to
single configuration Hartree-Fock calculations \citep{yuk-pot-rf}.

In contrast to the chemical picture, plagued by divergent partition
functions, the physical picture has the power to avoid them altogether.
Partition functions of bound clusters of particles ({\eg} atoms and ions)
are divergent in the
Saha approach, but has a compensating divergent scattering state part
in the physical picture \citep{ebeling:theory-b-states,fr:EOS1}.
A major advantage of the physical picture is that it incorporates this
compensation at the outset. A further advantage is that no assumptions about
energy-level shifts have to be made; it follows
from the formalism that there are none.

As a result, the Boltzmann sum appearing in the atomic (ionic) free energy is
replaced by
the so-called Planck-Larkin partition function (PLPF), given by
\citep{ebeling:theory-b-states,kraeft:quant-stat-particles,fr:PLPF}
\begin{equation}
	\nonumber
	{\rm PLPF} = \sum_{is} g_{is}
				 \left[ \exp\left(-\frac{E_{is}}{\kB T}\right)\ - \ 1 \ + \
				 \frac{E_{is}}{\kB T} \right] \ .
\end{equation}
%
The PLPF is convergent without additional cut-off criteria as are required in
the chemical picture. We stress, however, that despite its name the PLPF is
not a partition function, but merely an auxiliary term in a virial coefficient
\citep[see, {\eg},][]{wd_PL}.

The major disadvantage of the physical picture, is its formulation in terms of
density- or activity-expansions.
Expansions that first of all are very cumbersome to
carry out, which means that  so far only terms up to $\frac{5}{2}$ in density
have been evaluated \citep{al92,al94,al95}.
Second, the
slow convergence of the problem, means that even this extraordinary
accomplishment has a rather limited range of validity. The chemical
picture, on the other hand, do not need to rely on expansions, and 
complicated expressions, possibly with the correct asymptotic behavior,
can be used freely.

\subsection{The Coulomb correction}
\label{Coulomb-corr}

The Coulomb correction, that is, the consequence of an overall
attractive binding force of a neutral plasma deserves close attention,
because it describes the main truly non-ideal effect under conditions as
found in the interior of normal stars.
%
Already in a number of early papers
\citep[{\protect{\eg~}}][]{be80,ul82,ur83,sng83,sng84} it was suggested that
improvements in the equation
of state, especially the inclusion of a Coulomb correction, could reduce
discrepancies between computed and observed $p$-modes in the Sun. 
Responding to this, \citet{jcd-Nature:osc-eos}, showed that the MHD
equation of state indeed improved the agreement with helioseismology.
That the largest change was caused by the Coulomb correction was
not immediately clear, since the MHD equation of state also
incorporates other improvements over previous work.

From early comparisons between the MHD and OPAL equations of state
\citep{eoscmp1},
it turned out, rather surprisingly, that the net effect of the other
major improvement, the influence of hydrogen and helium bound states on
thermodynamic quantities, became to a large degree eclipsed beneath the
influence of the Coulomb-term. In the solar hydrogen and helium ionization
zones the Coulomb-term is the dominant correction to the ionizing perfect gas.
This discovery led to an upgrade of the simple, but astrophysically
useful \citet{EFF} (EFF) equation of state
through the inclusion of the Coulomb interaction term (CEFF)
\citep[see][]{cd91,jcd-wd:osc-eos}.

The leading-order Coulomb correction is given by the
Debye-H\"uckel (DH) theory, which replaces the
long-range Coulomb potential with a screened potential, as outlined below.

\subsubsection{The Debye-H\"uckel approximation}
\label{DHapprox}

The \citet{Deb:Huck} theory of electrolytes,
describes polarization in liquid solutions of electrons and positive ions. 
This description also applies to ionizing gases. Assuming the particles can move
freely, the electrons will congregate around the ions, and the ions will repel
each other due to their charges.
With their smaller mass and higher speeds, the paths of electrons are
deflected by the ions increasing the chance of finding an electron closer
to an ion.
This screening by the electrons decreases the repulsion between the ions,
acting as an overall attractive force in the plasma.

The fundamental assumption of Debye and H\"uckel is that of statistical
equilibrium, according to which the local density of particles of type $j$
(including electrons) immersed in a potential $\psi$ around an ion, $i$,
can be expressed as
\begin{equation}
	n_j({\bvec r}_i) = \langle n_j\rangle \exp(-Z_je\psi({\bvec r}_i)/\kB T)\ ,
\end{equation}
where $Z_je$ and $\langle n_j\rangle$ are the charge and mean density of
the particles and $n_j({\bvec r})$ are the perturbed densities.
$\psi({\bvec r})$ is the plasma-potential or the effective (screened)
inter-particle potential.
Over-all charge neutrality dictates that
\begin{equation}\label{no-net-charge}
	\sum_j\langle n_j\rangle Z_j = 0\quad\Leftrightarrow\quad
	 \langle n_{\rm e}\rangle  = \sum_{j \neq {\rm e}}\langle n_j\rangle Z_j\ .
\end{equation}
With these perturbed densities, the corresponding charge density is
\begin{equation}\label{charge-density}
	\nonumber
	\rho({\bvec r}_i) = \sum_jZ_je\langle n_j\rangle
					e^{-Z_je\psi({\bvec r}_i)/\kB T} + Z_ie\delta({\bvec r}_i)
\end{equation}
resulting in the Poisson equation
\begin{equation}\label{pois:boltz}
	\nabla^2\psi({\bvec r}_i) =
	-4\pi e\left[\sum_jZ_j\langle n_j\rangle
		e^{-Z_je\psi({\bvec r}_i)/\kB T} + Z_i\delta({\bvec r}_i)\right]\ .
\end{equation}
To make Eq.\
(\ref{pois:boltz}) more tractable, the exponential is expanded in a power
series.
The most critical of Debye's approximations is to retain 
only terms up to first order. The zero-order term
is the net-charge, Eq.\ (\ref{no-net-charge}). Solving Eq.\ (\ref{pois:boltz})
with the remaining first-order terms results in a screened Coulomb
potential---the {\em Debye-H\"uckel potential}
\begin{equation}\label{debye_pot}
	\psi(r) = \frac{Ze}{r}e^{-r/\lambda_{\rm DH}}\ ,
\end{equation}
where $\lambda_{\rm DH}$ is the Debye-length 
\begin{equation}\label{DHcl}
	\lambda_{\rm DH}^{-2} = \frac{4\pi e^2}{\kB T}\sum_i Z_i^2 n_i\ .
\end{equation}
The approximation of disregarding higher order terms affects the low temperature
and high density region where the inter-particle interactions becomes too
large to be described by just the first order term. This is a manifestation of
the problems with the classical, long-range part of the Coulomb field in
a plasma.

Investigations taking the physical picture point of view indicate that
the original potential defined in (\ref{pois:boltz}), is a good choice for
a plasma potential \citep{fr:EOS2}, and only the truncation of the exponential
resulting in the Debye-H\"uckel potential is of limited validity \citep{fr:EOS}

At high densities the effect is in fact overestimated by using the
Debye-H\"uckel potential (\ref{debye_pot}).
The relative Coulomb pressure in the Debye-H\"uckel theory, expressed
in terms of the coupling parameter,
$p_{\rm DH}/(n\kB T) = -\Gamma^{3/2}/\sqrt{12}$,
is a negative contribution to the pressure.
At very high densities, the over-estimation of the Coulomb pressure can be so
severe as to result in a negative total pressure.
The negative pressure differences seen in the comparison plots in Sects.\
\ref{EOSscape} and \ref{EOSsun}, suggests that the amplitude of the Coulomb
pressure is larger in OPAL than in MHD. This statement is true when the
$\tau$-correction, mentioned below, is applied to the MHD EOS.

To get a feeling for the behavior of the Coulomb pressure, we use the perfect
gas law to obtain the approximate expression
\begin{equation}\label{Lambda-approx}
	\Gamma \propto R^{1/3}\mu^{-1/3}\langle Z^2\rangle^{1/3}\ ,
\end{equation}
where $\mu$ is the mean-molecular weight.
This leads us to anticipate differences between OPAL and MHD, stemming from
different treatments of the plasma interactions, to increase with
$R$, and that such differences will be somewhat reduced when we mix in
helium and metals.

\subsubsection{The $\tau$ correction in DH theory}
\label{tau}

As they were investigating electrolytic solutions of molecules under
terrestrial conditions, it
was natural for Debye and H\"uckel to consider electrolytes made up of hard
spheres. Assuming there is a {\em distance of closest approach},
$r_{\rm min}$ to the ion, Eq.\ (\ref{debye_pot}) is modified to
\begin{equation}\label{debye_pot2}
	\psi(r) = \frac{Ze}{1+r_{\rm min}/\lambda_{\rm DH}}
	\frac{e^{-(r-r_{\rm min})/\lambda_{\rm DH}}}{r}\ ,
\end{equation}
for $r\ge r_{\rm min}$
and constant, $\psi(r_{\rm min})$, inside, removing the short range divergence.
To obtain the free energy, we apply the so-called {\em recharging} procedure
detailed in \citet{electrolyt} to Eq.\ (\ref{debye_pot2}), and  
get the result without $r_{\rm min}$, multiplied by the factor
\begin{equation}\label{taueq}
	\tau(x) = 3[\ln(1+x)-x+\frac{1}{2}x^2]x^{-3}\ ,
\end{equation}
where $x = r_{\rm min}/\lambda_{\rm DH}$.
In short, the recharging procedure consists of varying all charges in
the potential and integrating from zero to full charge.
Equation (\ref{taueq}) is the analytical result of this integration and
is based on the
assumption that $r_{\rm min}$ is independent of the charge of
any particles.
The $\tau$-factor goes from one to zero as $x$ increase,
reducing the Coulomb pressure which
was overestimated before. With the $\tau$ correction we can avoid the
negative pressures mentioned above.

\citet{tau} proposed to use
\begin{equation}\label{rminMHD}
	r_{\rm min} = \langle Z\rangle
e^2\left[\kB T\frac{F_{3/2}(\eta_{\rm e})}{F_{1/2}(\eta_{\rm e})}\right]^{-1}\ ,
\end{equation}
for stellar plasmas, and it was later used in the MHD EOS but not in OPAL.
This choice of $r_{\rm min}$ is merely the distance of equipartition between
thermal and potential
energy of electrons approaching ions. Since the charges are opposite there are,
however, no classical limits to their approach.
Also notice that since this choice of
$r_{\rm min}$ depends explicitly on charge, the recharging procedure will
result in a different form of $\tau$.

A thorough and critical review of the Debye-H\"uckel theory can be found in
\citet{electrolyt}, Chp. IX, and a very clear presentation is found in
\citet{sse:deb},
though the latter does not mention $\tau$.

\subsubsection{Other higher-order Coulomb corrections}
\label{HigherOrder}

Obviously, the $\tau$ correction is just one particular higher-order
Coulomb correction. We can use it as a model for developing
more general expressions,
by allowing some liberty in the choice of $r_{\rm min}$.
Let us begin by asking 
about the distance of closest approach for quantum-mechanical
electrons.
Heisenberg's uncertainty relation puts firm limits on how localized particle
can be --- it is smeared out over a volume the size of a 
de Broglie wavelength $\lambda=\hbar/p$. This de-localization eliminates the
infinite charge densities associated with classical point-particles, and hence
the short-range divergence of the Coulomb potential.

Based on that, we can tentatively suggest a distance of closest approach
which is the combined radii of the electron and ion:
$\frac{1}{2}\lambda_{\rm e} + \frac{1}{2}\langle\lambda_{\rm ion}\rangle$.
The diffraction parameter, $\gamma_{ij}$, between two particles $i$ and $j$,
emerging from a more careful quantum-mechanical analysis implies the use of
the de Broglie wavelength in relative coordinates
\begin{equation}\label{rmin-diff}
	r_{\rm min} = 
        \lambdabar_{\rm ij} = (\hbar^2/2\mu_{\rm ij}\kB T)^{1/2}
		\propto T^{-1/2}\ ,
\end{equation}
where $\mu_{ij}$ is the reduced mass.
Comparing the $\tau$-function with the quantum diffraction modification in
Fig.\ 5 of \citep{fr:EOS}, we see a similarity in the functional form.
The asymptotic behavior differs though: $\tau(x)\rightarrow x^{-1}$ for
$x\rightarrow \infty$ in the hard-sphere model, whereas quantum diffraction
goes as $x^{-1/2}$.
The two functions are very close up to $x\simeq 1$, though, 
suggesting that preliminary investigations of quantum diffraction effects in
the MHD EOS could be carried out by means of the $\tau$-function and a new
$r_{\rm min}$ as given by Eq.\ (\ref{rmin-diff}).

Dividing Eq.\ (\ref{rmin-diff}) by $\lambda_{\rm DH}$, we find that
the correction is now a function of $\varrho$ only. That is, going
from a hard-sphere model of interactions, to including quantum
diffraction, the factor alleviating the short-range divergence of the Coulomb
potential becomes a function of $\varrho$ instead of $R$.

Abandoning the hard-sphere ion correction for the benefit of quantum
diffraction, still leaves us with only the first term of the Coulomb
interactions.
Could the higher order terms be represented by $\tau$ in some form?
It turns out that $\tau$ would only
fit in a very limited range,
and it would be more fruitful to use proper expressions.
The present
analysis however, shows that the effect of including higher-order
Coulomb terms, is
smaller than has previously been estimated by the MHD EOS. It therefore might
be a better approximation to leave them out for at least the solar case.
As shown in Fig.~13 of \citet{Qmhd}, the coupling parameter attains appreciable
values in the outer layers of the Sun and higher-order Coulomb terms will
most likely cause better agreement with helioseismology.
\section{The EOS landscape in $\varrho$ and $T$}
\label{EOSscape}
For this comparison, we have 
computed MHD EOS tables with exactly the same
$\varrho/T$-grid points as the OPAL-tables \citep{rogers:OPAL-EOS},
to ensure that 
the equation-of-state comparison is not influenced by interpolation errors.
We do actually use the respective interpolation routines to access the
table-values,
but by interpolating on the exact grid-points for identical mixtures, we should
not lose precision in the process.

\begin{deluxetable}{lrrrr}
\tablecolumns{5}
\tablewidth{0pt}
\tablecaption{Chemical mixtures 2 and 3\label{mixes} (see text)}
\tablehead{
\colhead{element} & \colhead{$X_i(\%)$} & \colhead{[$N_i/N_{\rm H}$]}
				  & \colhead{$X_i(\%)$} & \colhead{[$N_i/N_{\rm H}$]}}
\startdata
~~H  & 80.00 &  0.00000 & 80.00    &  0.00000 \\
~~He & 20.00 & -1.20098 & 16.00    & -1.29789 \\
~~C  &  0.00 &   ---~~~ & 0.762643 & -3.09693 \\
~~N  &  0.00 &   ---~~~ & 0.223398 & -3.69693 \\
~~O  &  0.00 &   ---~~~ & 2.171950 & -2.76693 \\
~~Ne &  0.00 &   ---~~~ & 0.842053 & -3.27923 \\
\enddata
\end{deluxetable}
We compare
tables with three different chemical mixtures, successively
adding more elements to the plasma:
Mix~1 is pure hydrogen, Mix~2 a hydrogen-helium mixture and Mix~3 is a
6-element mixture that, besides hydrogen and helium, also includes
C, N, O and Ne.
In Table\ \ref{mixes} we
list the exact mixtures, both by mass abundance, $X_i$, of 
chemical element, $i$, and
as logarithmic number fractions relative to hydrogen [$N_i/N_H$]. 
The choice of mixtures is that of the currently available OPAL-tables,
to avoid interpolations in $X$ and $Z$.
In the comparisons of this section, we have omitted the
radiative contributions.

The MHD equation of state now includes relativistically degenerate electrons,
\citep{GDZ:rel-e-MHD} as do the new version of OPAL \citep{rogers:newOPAL}.
This, of course, is significant for stellar modeling and important for
helioseismic investigations of the Solar radiative zone \citep{elliott:RelDegSun}.
For the present paper, however, it is irrelevant due to the lack of
controversy on the subject, and we will therefore limit ourselves to
dealing with non-relativistic electrons.

All plots of differences in this paper present absolute differences.
Since the absolute quantities span less
than an order of magnitude and as they have quite
complicated behaviors, we found that normalizing the differences would 
confuse more than illuminate. The solar track (also
presented in Sect.\ \ref{EOSsun}) overlaid on
the surface plots is 
not hidden behind the surface, so as to give an idea of the behavior in
otherwise obscured regions.

While the MHD tables and the pure-hydrogen OPAL table have the same
resolution, Mix~2 and Mix~3 OPAL tables have three times higher
resolution both in $T$ and $\varrho$. 
This can only affect the comparisons of the
solar Mix~2 and 3 cases, Sect.\ \ref{EOSsun}, where it might introduce some
extra interpolation-wiggles in the OPAL-MHD differences.
The table comparisons are all done on the low
resolution grid.

For the case of pure hydrogen (Mix~1) 
we plot the logarithmic absolute pressure, but for
the other mixtures we plot the logarithm of a reduced pressure, $P/(\varrho T)$,
to make it easier to identify 
non-ideal effects and the location of ionization zones. This choice will
of course not affect the differences of the logarithms.

Apart from the actual pressure we also investigate the three derivatives
\begin{equation}
  \chi_\varrho= \dxdycz{\ln P}{\ln\varrho}{T}\,,\quad
  \chi_T      = \dxdycz{\ln P}{\ln T}     {\varrho}\,,{\rm and}\quad
  \gamma_1    = \dxdycz{\ln P}{\ln\varrho}{S}\ ,
\end{equation}
where $\gamma_1$ is the adiabatic derivative often called $\Gamma_1$.
These three derivatives form a complete set and fully describe the
equation of state.

\subsection{Pure hydrogen}
\label{TabH}
We start with the simplest mixture, that is, pure hydrogen (Mix~1).
The case of hydrogen is, however, far from simple, not the least because
of its negative ion and molecular species.
All in all five species of hydrogen: H, H$^+$, H$^-$, H$_2$ and H$^+_2$ are 
included in both EOS.

The number of negative hydrogen 
ions does never exceed a few parts in a thousand compared to
the other hydrogen
species. Already at moderate temperature, 
they dissociate into hydrogen atoms. Despite its low
abundance, H$^-$ does have an
impact on the electron balance since it is the only (significant)
electron sink. The heavy elements with their low abundances are most
affected by this.
Apart from this indirect effect on the heavy elements,
the most important feature of the H$^-$-ion is of course its bound-free and
free-free opacity, which is 
the primary source of opacity in atmospheres of G, K
and M stars.

The positive and neutral hydrogen molecules
can be seen in the low-temperature-high-density corner of the
tables, where their abundance reaches up to 28\%
of hydrogen, by mass. At slightly lower
densities, which is of greater astrophysical interest, these molecules only 
become important at temperatures below those considered here.

\begin{figure}
\plotone{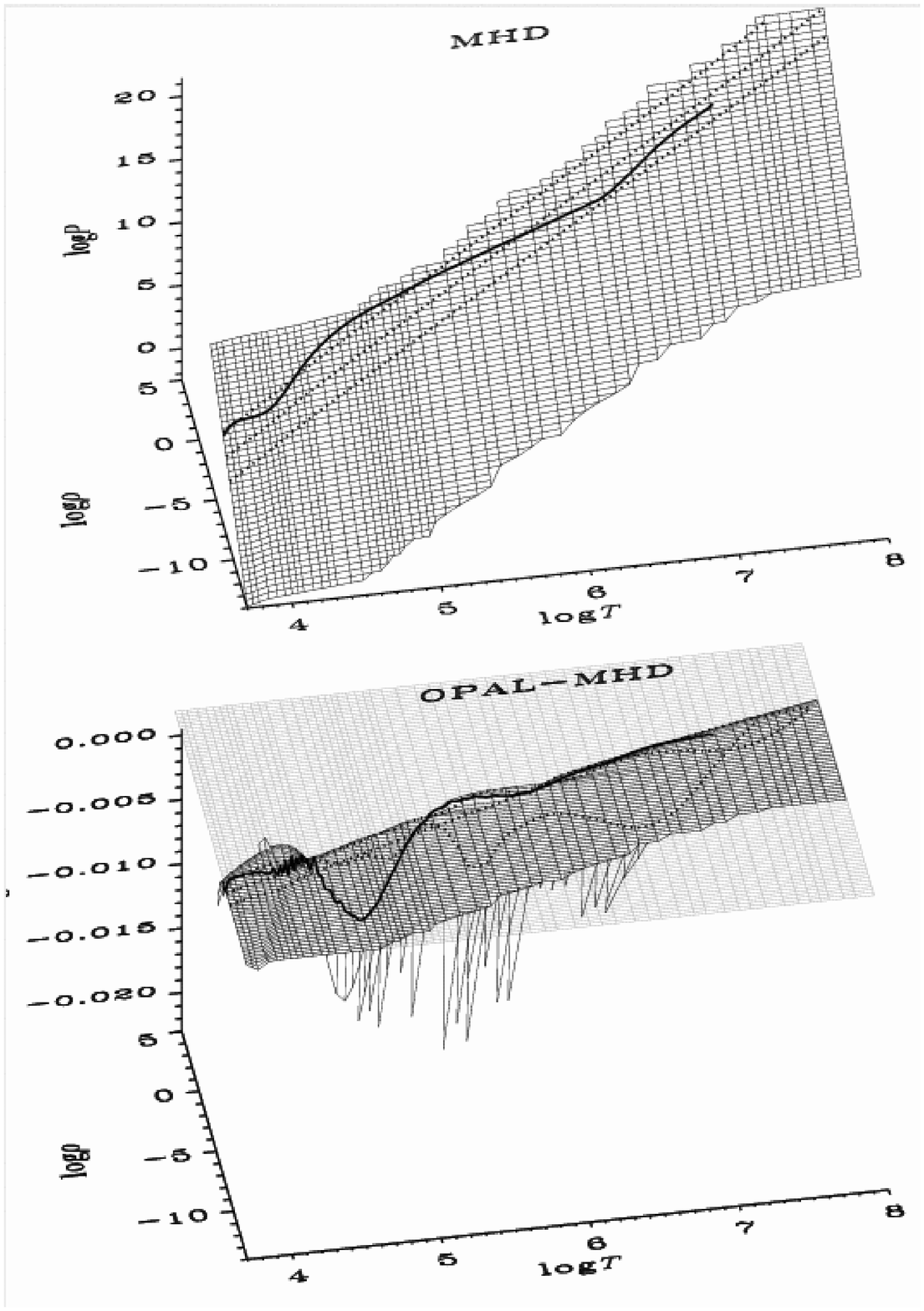}
\caption{Comparison of log$_{10}P$ in the two pure hydrogen tables. The upper
	panel shows the absolute value from the MHD EOS and the lower panel shows
	the difference; OPAL minus MHD. The strange boundaries of the surface simply
	reflects the shape of the tables. We also overlay the solar track from
	Sect.\ \ref{EOSsun} for comparison. On this plot alone we also show iso-$R$
	tracks (dotted lines) for log$_{10}R = -2, -1, 0$, going from low to high
	densities.\label{table_H_norad1}}
\end{figure}

\begin{figure}
\plotone{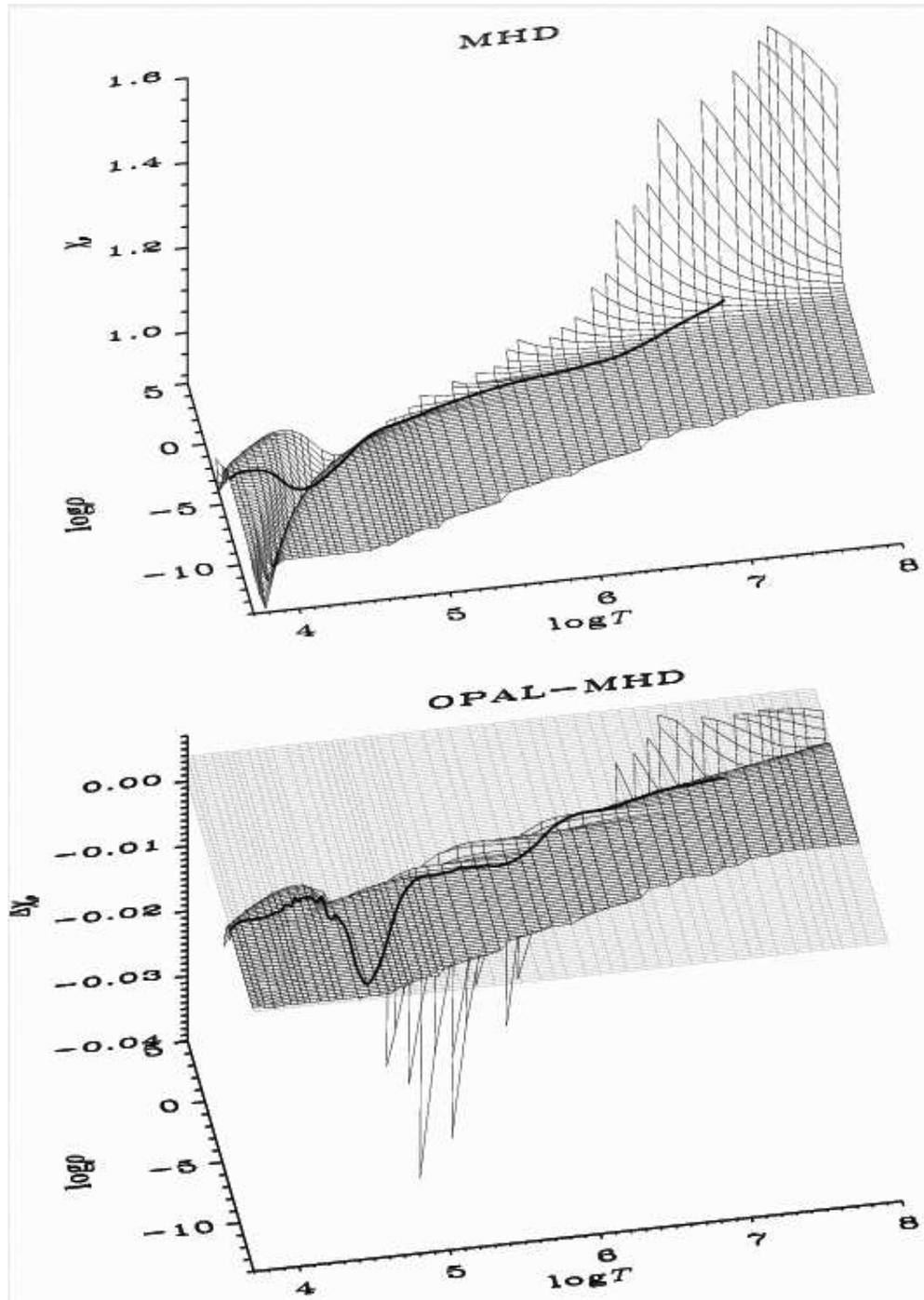}
\caption{The logarithmic pressure derivative with respect to density
	$\chi_\varrho=(\partial\ln P/\partial\ln\varrho)_T$ for pure hydrogen
	in the upper panel, and its differences (OPAL minus MHD) in the lower
	panel.\label{table_H_norad3}}
\end{figure}

\begin{figure}
\plotone{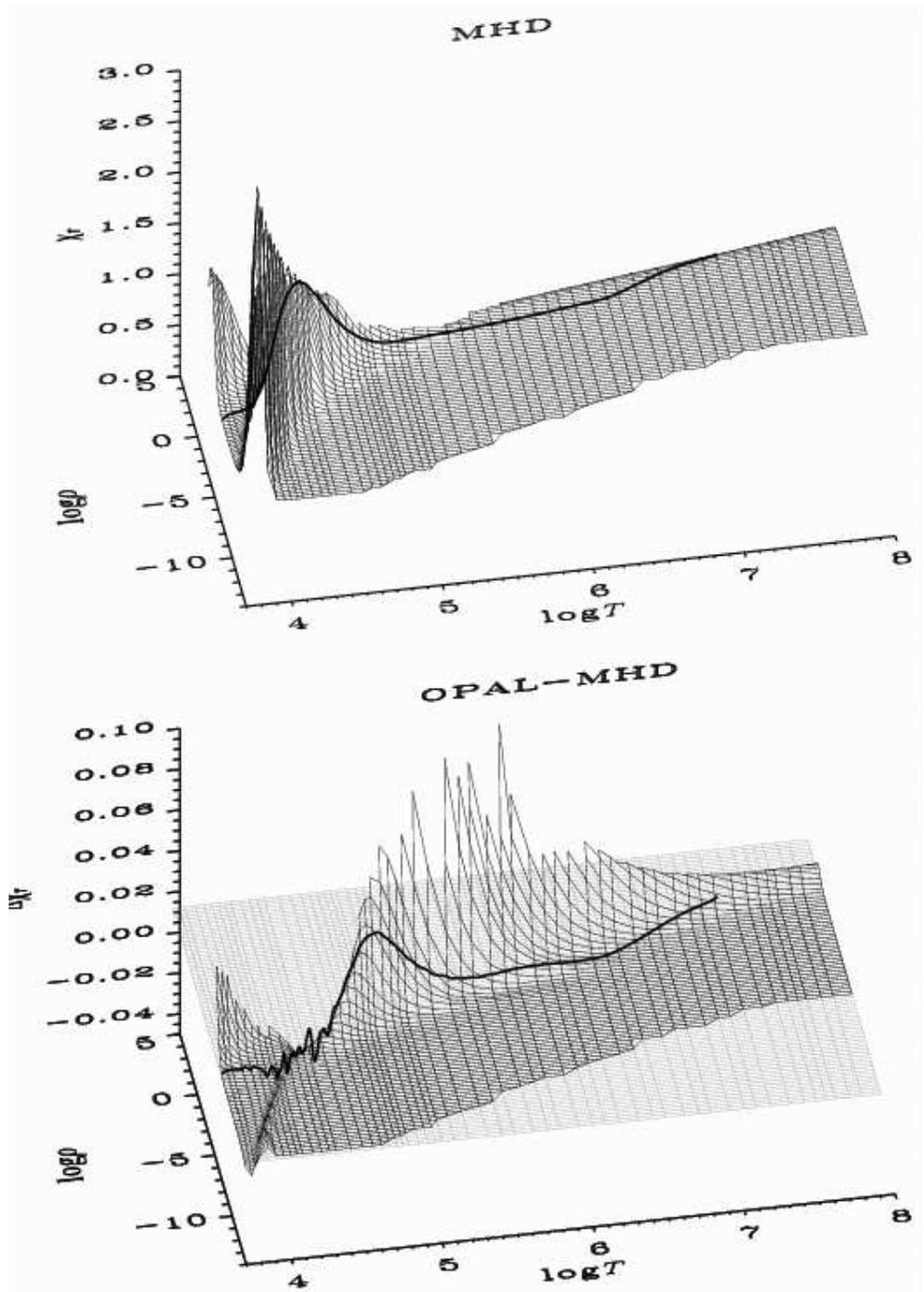}
\caption{The logarithmic pressure derivative with respect to temperature
	$\chi_T=(\partial\ln P/\partial\ln T)_\varrho$ for pure hydrogen
	in the upper panel, and its differences (OPAL minus MHD) in the lower
	panel.\label{table_H_norad4}}
\end{figure}

\begin{figure}
\plotone{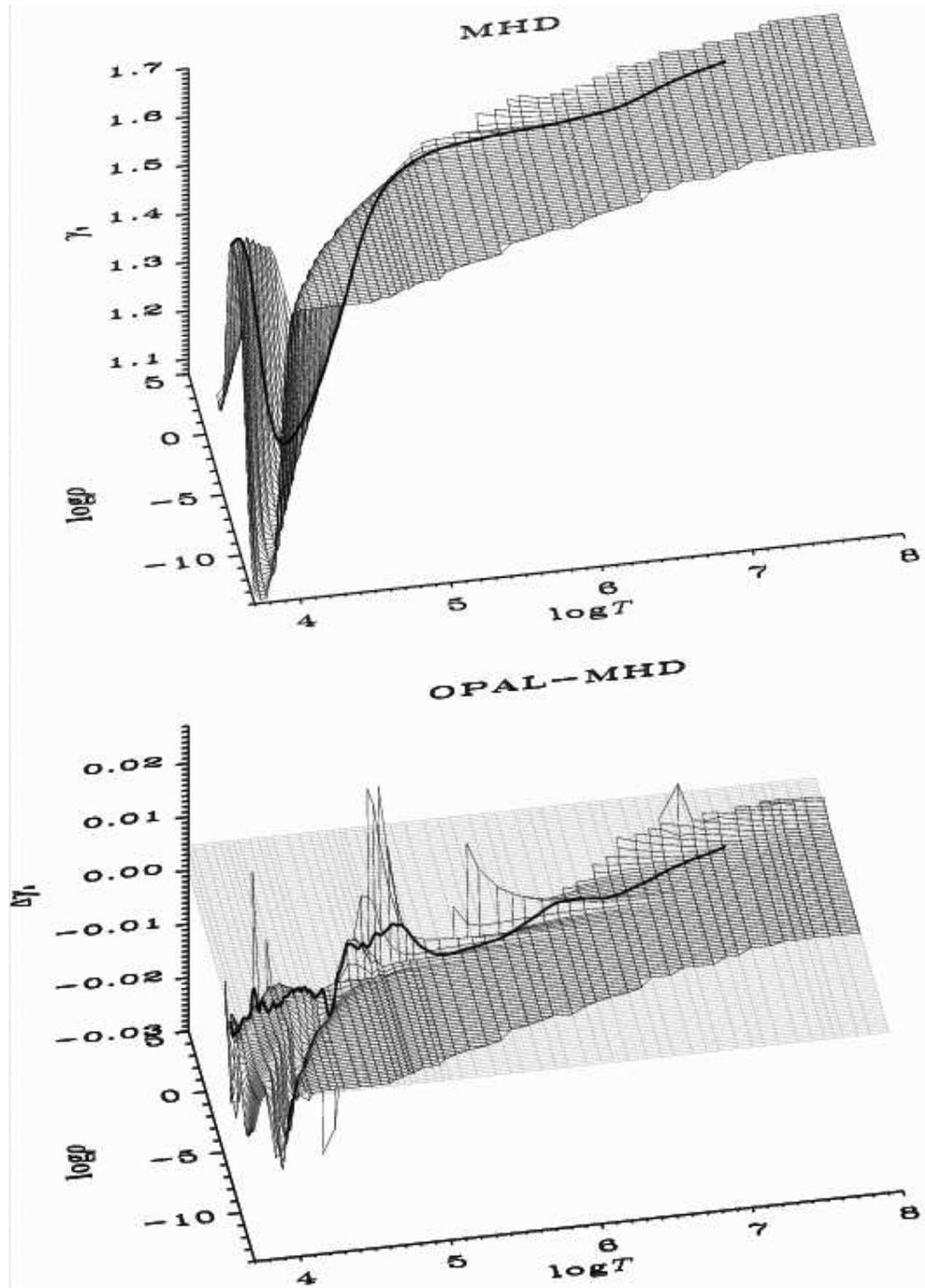}
\caption{The adiabatic logarithmic pressure derivative with respect to density
	$\gamma_1=(\partial\ln P/\partial\ln\varrho)_S$ for pure hydrogen
	in the upper panel, and its differences (OPAL minus MHD) in the lower
	panel.\label{table_H_norad8}}
\end{figure}

The most important feature in the hydrogen-EOS landscapes of
Figs.\ \ref{table_H_norad1}--\ref{table_H_norad8}~is, by far, the ionization
(from atom to positive ion), seen as a curved rift in all the
derivatives. It is hardly visible in the surface-plot of the full pressure 
(Fig.\ \ref{table_H_norad1}), but becomes 
obvious in those of the reduced pressure (Figs.\
\ref{table_HHe_norad1} and \ref{table_H-Ne_norad1}).

The OPAL-team introduced the quantity
\begin{equation}\label{R}
	R = T_6^3\varrho^{-1}\ ,
\end{equation}
where $T_6 = T/10^6$, as a convenient quantity to describe the approximate
$\varrho-T$-stratification of many stars. This is clearly seen in the upper
panel of Fig.\ \ref{table_H_norad1} where we also plotted three iso-$R$ tracks,
bracketing the solar track.
Since the full pressure surface is so close to a plane, this plot conviniently
shows the range and borders of the tabels. When interpreting the surface plots
in the following sections, keep in mind this non-rectangular shape of the
tables.
In the lower panel of Fig.\ \ref{table_H_norad1} these iso-$R$ tracks also
bracket the main feature in the differences:
A sharply rising ridge, bell-shaped in
log$T$ and centered around log$T=5.5$, approximately aligned with log$R=0$.
This ridge is a signature of differences in the pressure
ionization. The sign of $\Delta P$ in 
Fig.\ \ref{table_H_norad1} tells us that
MHD has a more abrupt pressure
ionization than the softer OPAL.
The reason for this difference is still not completely
clear. It might be related
to differences in the treatment of the
short-range suppression of the Coulomb forces,
as mentioned in Sect.\ \ref{tau} and \ref{HigherOrder}, or it could be a
result of differences in the mechanism of pressure ionization
\citep{iglesias:OPAL-OP-diff,basu:H-part-func,GDN:Z+partfnc-eff}.
\placefigure{table_H_norad1}

We now turn to the logarithmic pressure derivatives, $\chi_\varrho$ and
$\chi_T$,
displayed in Fig.\ \ref{table_H_norad3} and \ref{table_H_norad4}, respectively.
In both figures, the ionization zone is easily recognized as the canyon or ridge
starting in the low-temperature-low-density corner, slowly bending over
to follow the solar track and disappear at about log$T=6$.

In most of the $\varrho-T$ plane both derivatives are equal
to one reflecting that the gas is a perfect gas. 
In this region the differences are very small ({\ie} less than 0.03\%),
confirming that both the chemical and the physical picture converges
appropriately to the perfect gas case.

At low temperatures $\chi_\varrho$ and $\chi_T$ are dominated by temperature
ionization, which is about an order of magnitude more prominent in $\chi_T$
than in $\chi_\varrho$. This region is a fairly well known regime and here we
can directly compare
the two pictures. The differences are indeed small in this region, less than 1\%
and less than a tenth of the differences in the high-$R$ ridge.

The rise of $\chi_T$ in the low-$T$-high-$\varrho$
corner is due to H$_2$-molecules. About 28\% by mass, of the hydrogen atoms 
are bound in molecules in this region, but at higher densities they quickly 
dissociate. As beautifully illustrated by \citet{mhd2} in their Fig.~1, hydrogen
recombines fully to H$_2$ at large densities when assuming a Saha equilibrium.
This is obviously absurd (there is no room for molecules -- or atoms) and both
the MHD and OPAL EOS pressure dissociate hydrogen here, as would be expected
from a realistic EOS.
In MHD this is modeled in the same manner as
the pressure ionization of atoms, as explained below Eq.~(\ref{Zs}). This is
most likely a rough approximation, but due to the efficient ionization with
density, the details might be less important.
The fact that 
the differences increase while the absolute value decreases indicates that
MHD is pressure dissociating faster than OPAL.

The differences are again dominated by the sharp ridge at high $R$, but in
contrast to pressure (Fig.\ \ref{table_H_norad1}), 
the differences in $\chi_\varrho$ and $\chi_T$ return to zero for high
temperatures and densities. As for pressure, the
solar track falls over or climbs the $R$-edge in the middle of the ionization
zone, as it is traversing the iso-$R$ at log$R\simeq 0$.
\placefigure{table_H_norad3}

These high-$R$ differences occur in a region where there is competition
between the Coulomb terms and electron degeneracy. This makes the
interpretation much more difficult.
Two possible reasons are the previously mentioned short-range part of
the Coulomb interactions and the changes induced in 
the internal atomic states by the
dense, perturbing surroundings.

In the MHD EOS, all energy levels 
of internal states are assumed to be unaltered by the plasma
environment. That is, the effect 
of the perturbation by surrounding neutral and charged
particles on the internal state is restricted to a 
lowering of the occupation probability of the given state only.
In the OPAL EOS, the net result looks similar, but there the
relative stability of energy levels to perturbations is not
merely postulated but the result of in-situ calculations
of the Schr\"odinger or Dirac equation for each
configuration of nuclei and electrons, based on 
parameterized Yukawa potentials
\citep{yuk-pot-rf}, as mentioned in Sect.\ \ref{physEOS}.
\placefigure{table_H_norad4}

Looking at $\gamma_1$ in Fig.\ \ref{table_H_norad8} 
we immediately notice how well this
quantity displays the ionization zones while leaving out everything else.
This property is also reflected in the differences, which here are of
about the
same magnitude in the ionization zone as in the high-$R$ ridge.
The high-$R$ differences have also changed characteristics, 
changing sign periodically, while retaining the overall bell-shape
in log$T$ of the amplitude.
We mention, however, that 
at least some of this behavior might be due to the numerical
differentiation scheme used in the OPAL EOS (see Sect.\ \ref{diff} and
Fig.\ \ref{gamma1}).
\placefigure{table_H_norad8}
\subsection{Hydrogen and helium mixture}
\label{TabHHe}
\begin{figure}
\plotone{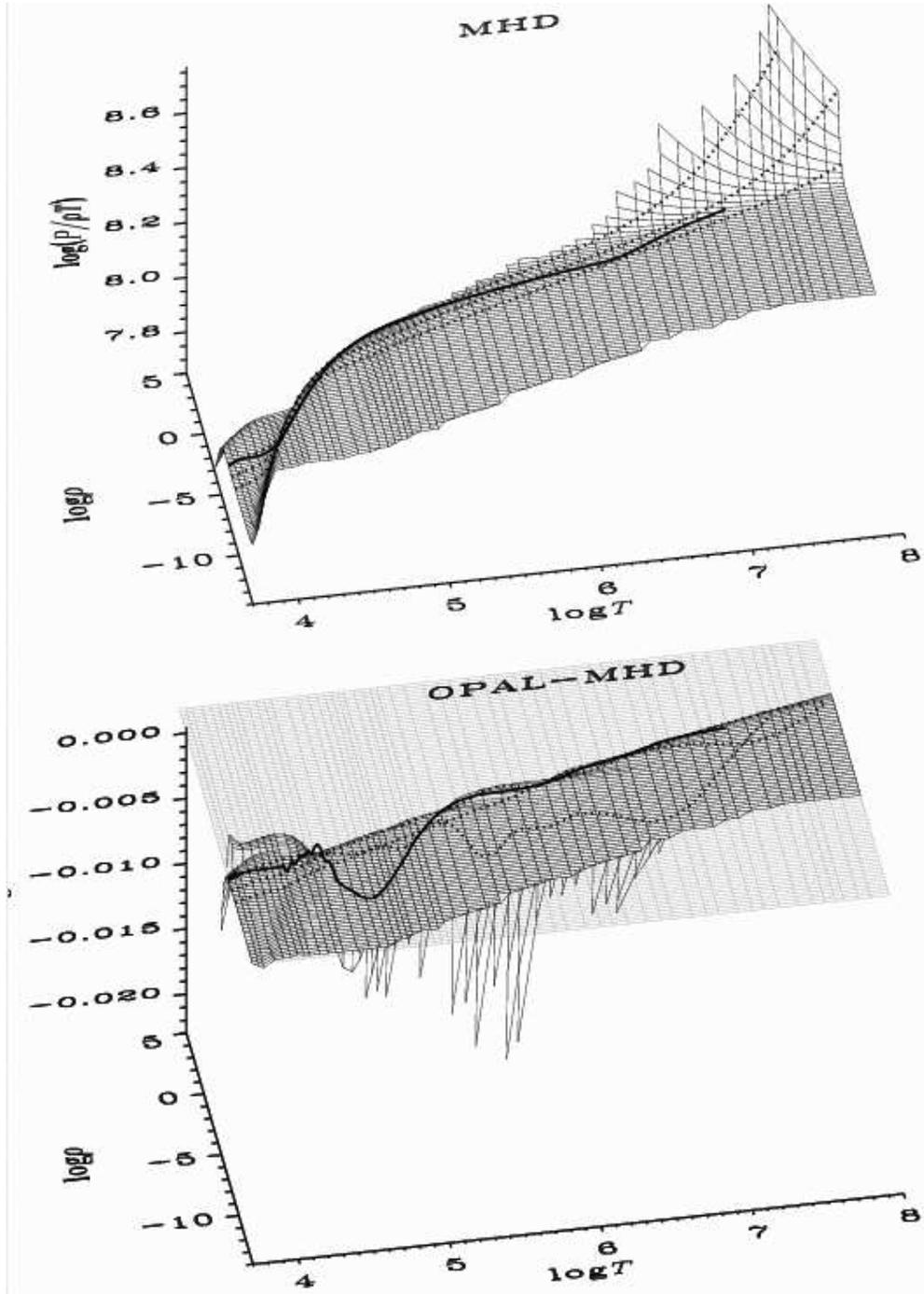}
\caption{The reduced pressure, $P/(\varrho T)$, for the H-He mixture
	in the upper panel, and its differences (OPAL minus MHD) in the lower
	panel.\label{table_HHe_norad1}}
\end{figure}

The effect of helium in the thermodynamical quantities is revealed by
the addition of 20\% helium and comparison with the pure-hydrogen case.
The first
thing we notice from Fig.\ \ref{table_HHe_norad1} is how well the 
reduced pressure
$P/(\varrho T)$ reveals all the dissociation and ionization zones (except
H$^-$); The H$_2$-formation in the low-$T$-high-$\varrho$ corner and the
prominent ionization of hydrogen together with the two He ionization zones,
the first eventually merging with the H ionization.
The effect of degenerate electrons is evident in the high-$T$-high-$\varrho$
corner.

We also notice another thing: while the pure
hydrogen OPAL-table was cutting the high-$\varrho$, low-$T$ corner, leaving a
little less for the comparison, the mixture OPAL-tables allow a full
comparison since they have the
same boundaries as the MHD tables. The slightly larger table reveals a new
feature in the differences. For pure hydrogen, the pressure difference drops
suddenly in the high-$\varrho$, low-$T$ corner, due to faster molecule 
formation in OPAL as compared to MHD.
But in the slightly larger tables used for the remainder of this section, this
difference suddenly goes to zero before it falls down the high-$R$ edge.
\placefigure{table_HHe_norad1}

In the pressure differences, one can just barely identify the first helium
ionization zone, whereas the second is too faint to be seen here. 
The high-$R$ differences
are a little smaller than for pure hydrogen, as anticipated from
Eq.\ (\ref{Lambda-approx}) and the discussion following it.
This can be most clearly seen by
comparing the dip in the hydrogen ionization zone.

\begin{figure}
\plotone{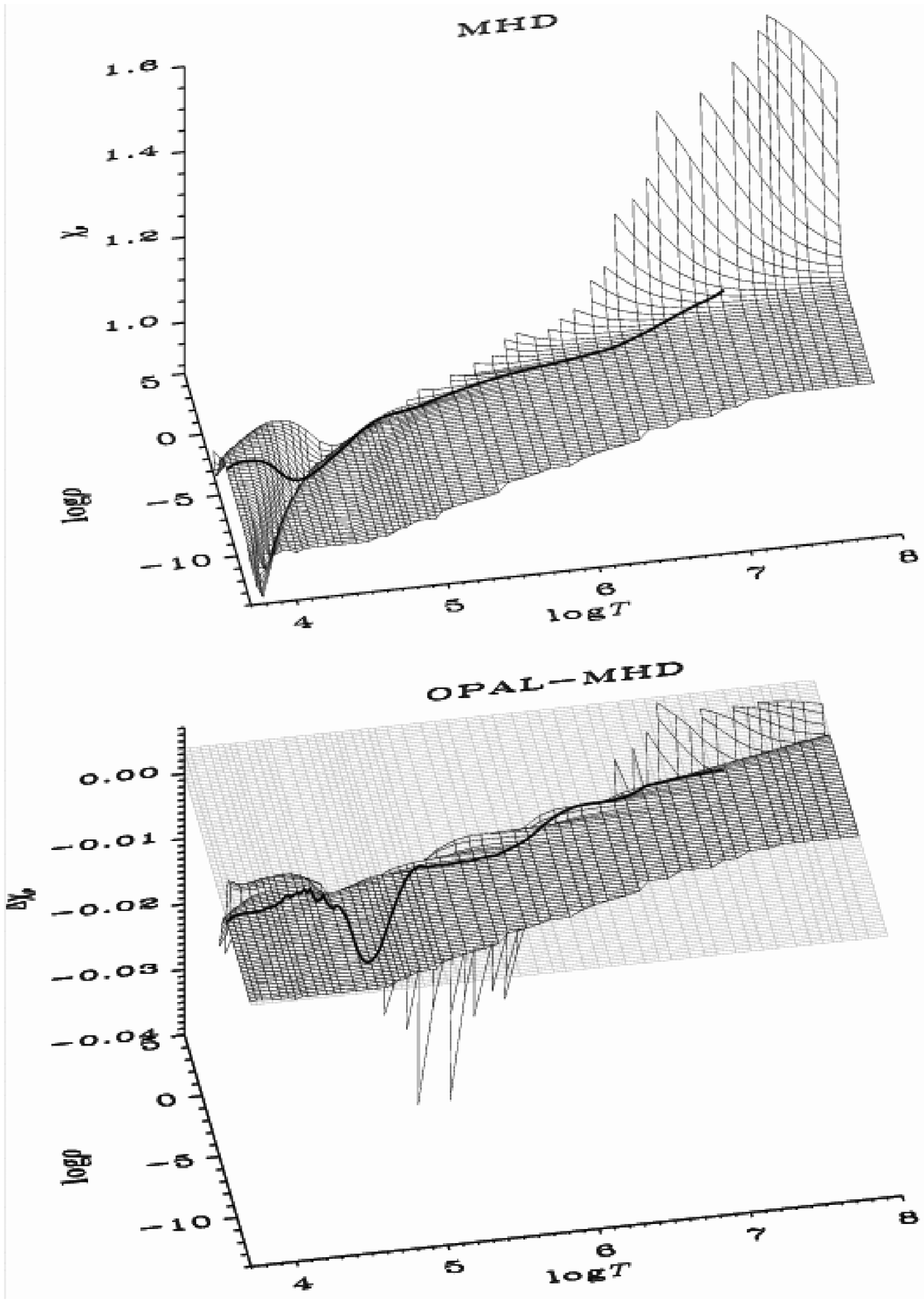}
\caption{$\chi_\varrho$, the logarithmic pressure derivative at constant temperature, for
	the H-He mixture in the upper panel, and its differences (OPAL minus MHD)
	in the lower panel.\label{table_HHe_norad3}}
\end{figure}

\begin{figure}
\plotone{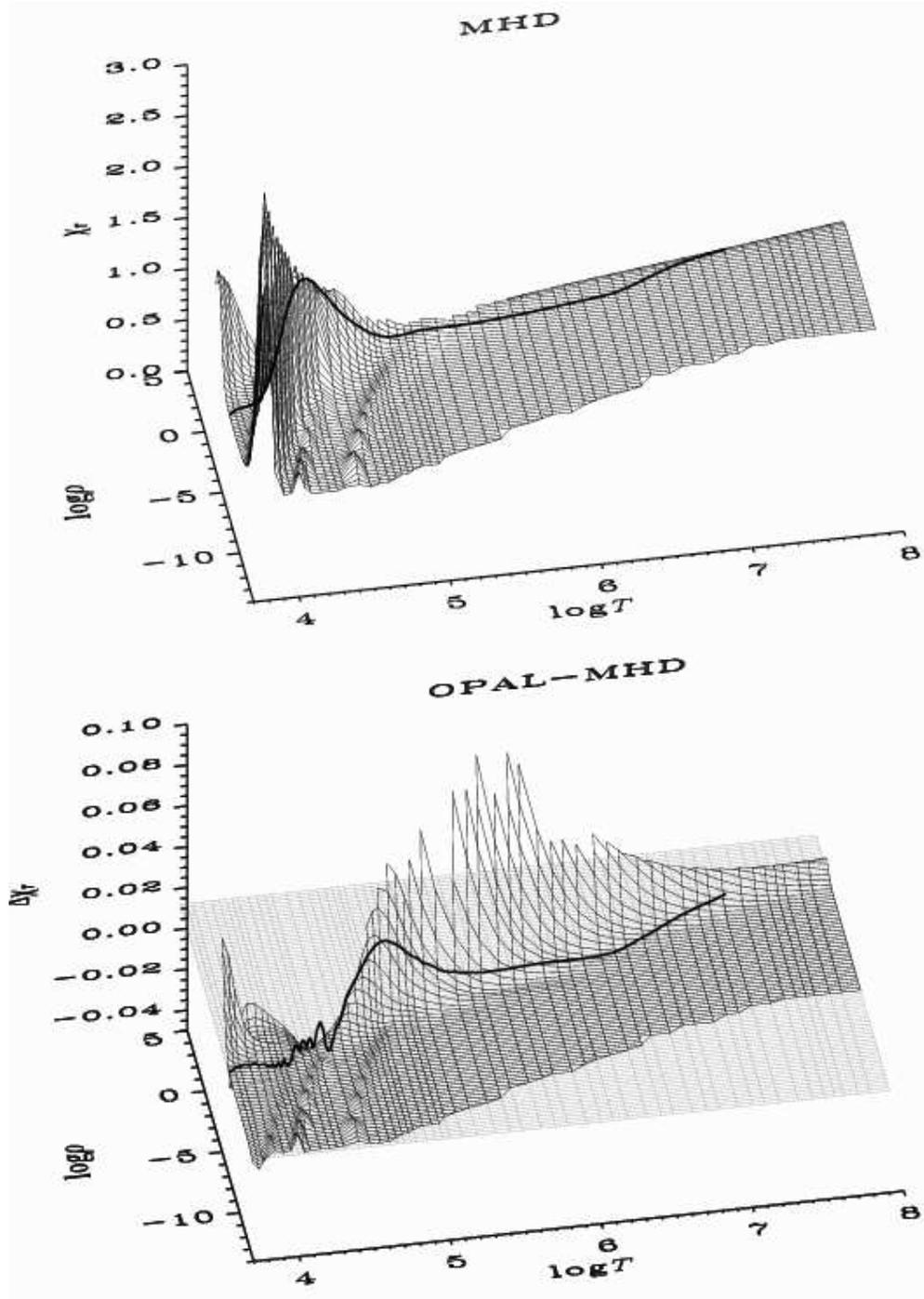}
\caption{$\chi_T$, the logarithmic pressure derivative at constant density, for
	the H-He mixture in the upper panel, and its differences (OPAL minus MHD)
	in the lower panel.\label{table_HHe_norad4}}
\end{figure}

The addition 
of helium is also evident in the logarithmic pressure derivatives in
Fig.\ \ref{table_HHe_norad3} and \ref{table_HHe_norad4}. 
First we see the deep rift (ridges in
Fig.\ \ref{table_HHe_norad4}) of
the hydrogen ionization zone. Then comes a small groove from the first
helium ionization zone, a groove which, when it widens and gets shallower at
higher densities, eventually merges with the hydrogen ionization zone, as
is the case for the solar track. 
Widely separated from the hydrogen and first helium ionization zones,
we find
the second helium ionization zone. It seems to disappear
at the low density edge of the table, but that is only so because the ridge
gets very sharp and is unresolved in temperature, at low densities.
Hotter stars, that is, stars shifted towards
lower $R$, will clearly exhibit three, 
more distinct ionization zones when compared with
the Sun.
\placefigure{table_HHe_norad3}

Apart from the two helium ionization zones, the differences in the pressure
derivatives are very similar to the pure hydrogen case. The high-$R$
differences are somewhat smaller though, as are the differences in the hydrogen
ionization zone. The differences in $\chi_\varrho$ also displays
a very small ripple along log$R\simeq -4$, which might be due
to differences in the differentiation technique (see Sect.\ \ref{diff}).
\placefigure{table_HHe_norad4}

From the differences in $\chi_T$ (Fig.\ \ref{table_HHe_norad4}), we see that the
absolute differences in the three ionization zones are just about the same. If
we instead compare the differences relative to the size of the respective
ionization ridges, we get 0.16\% and 3\% relative differences for the hydrogen
and helium ionization zones respectively. That is, MHD and OPAL have about 20
times better agreement on hydrogen than on helium.

\begin{figure}
\plotone{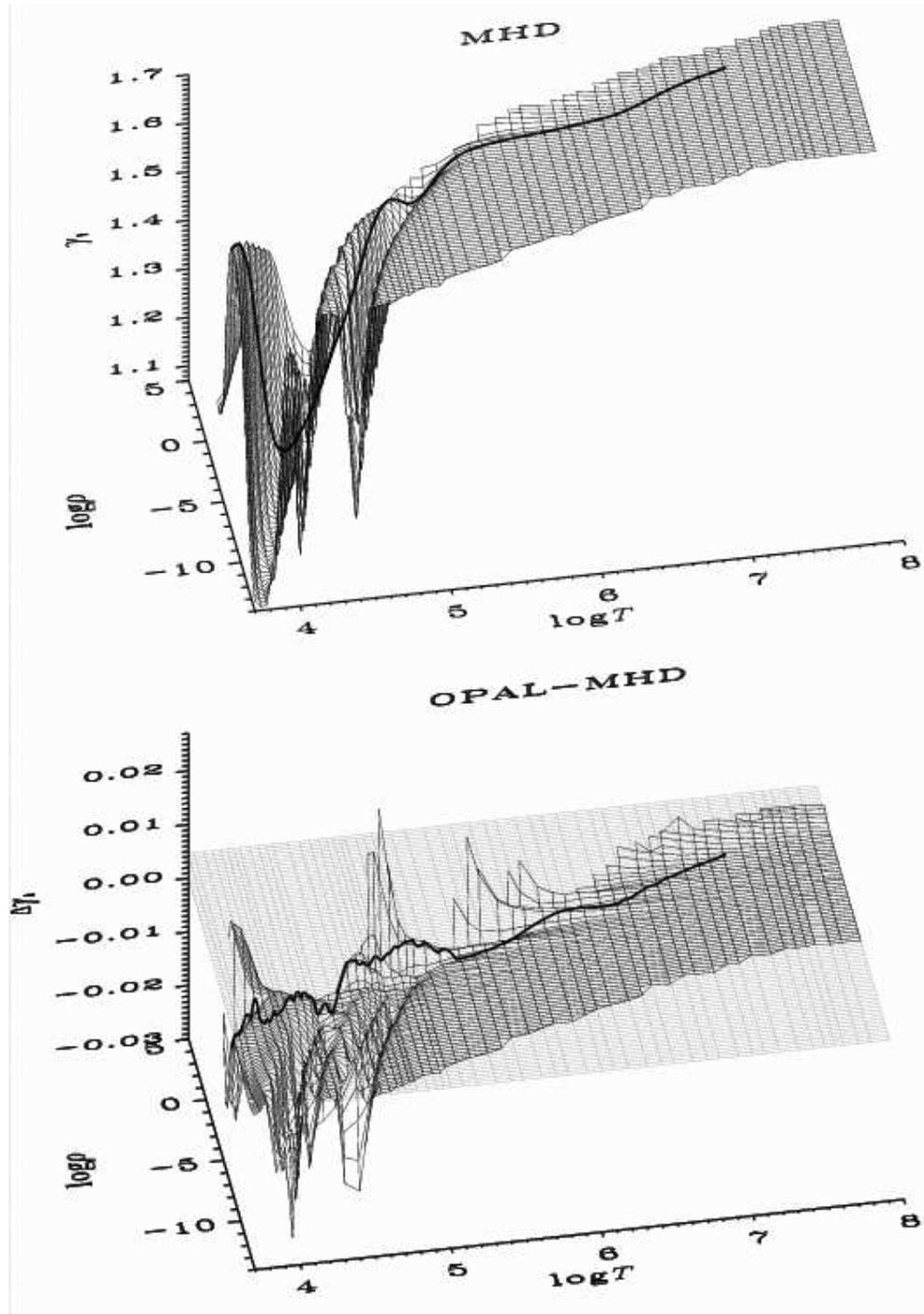}
\caption{$\gamma_1$ for the H-He mixture in the upper panel, and its differences
	(OPAL minus MHD) in the lower panel.\label{table_HHe_norad8}}
\end{figure}

In Fig.\ \ref{table_HHe_norad8}, $\gamma_1$ appears like what 
we would anticipate from the
pure hydrogen case in Fig.\ \ref{table_H_norad8}. 
The first helium ionization zone is
only visible at low densities, as it merges with the hydrogen ionization zone
shortly before the solar track is reached.
\placefigure{table_HHe_norad8}

The differences, however, exhibit a much more complicated
structure. Along each of the ionization zones, there is a deep
valley in the differences, and along the bottom of these valleys runs a very
sharp ridge, bringing the differences up to positive values. This is a clear
sign of a broad negative peak minus a sharp negative peak, meaning that MHD
temperature ionize faster than does OPAL. In the beginning of this section, we
found that MHD was also pressure ionizing faster than OPAL, so all in all OPAL
is the softer EOS of the two. The ridge-in-the-middle-of-the-valley picture is
also found in the pure hydrogen case (Fig.\ \ref{table_H_norad8}), 
but as the hydrogen
ionization zone is not fully covered at low densities, the low-$T$ side of
the valley is missing.
\subsection{H,He,C,N,O and Ne mixture}
\label{TabH-Ne}
\begin{figure}
\plotone{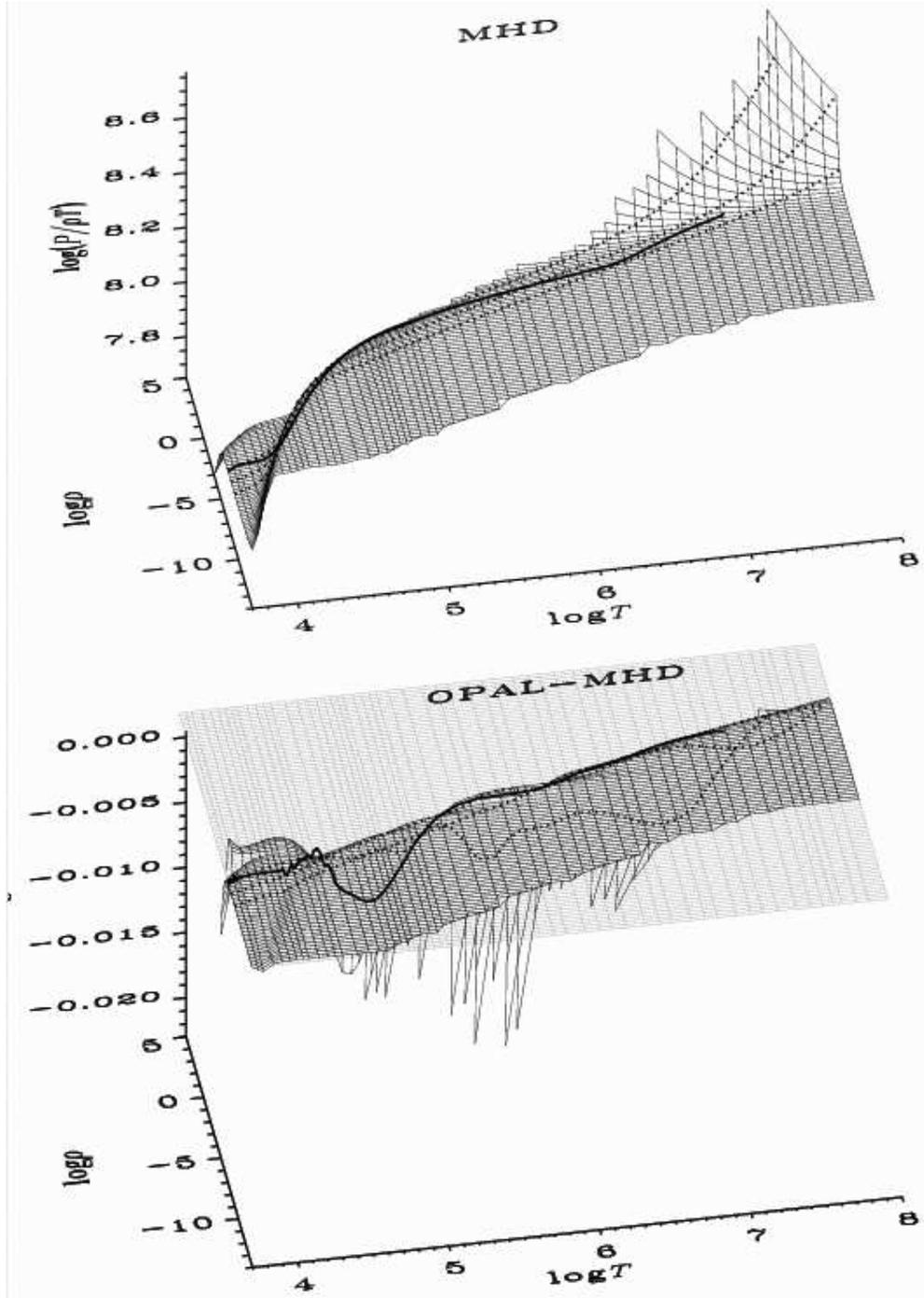}
\caption{Reduced pressure for mixture 3 ({\cf} Tab.\ \ref{mixes}) in the upper
	panel, and its differences (OPAL minus MHD) in the lower panel.
	\label{table_H-Ne_norad1}}
\end{figure}

\begin{figure}
\plotone{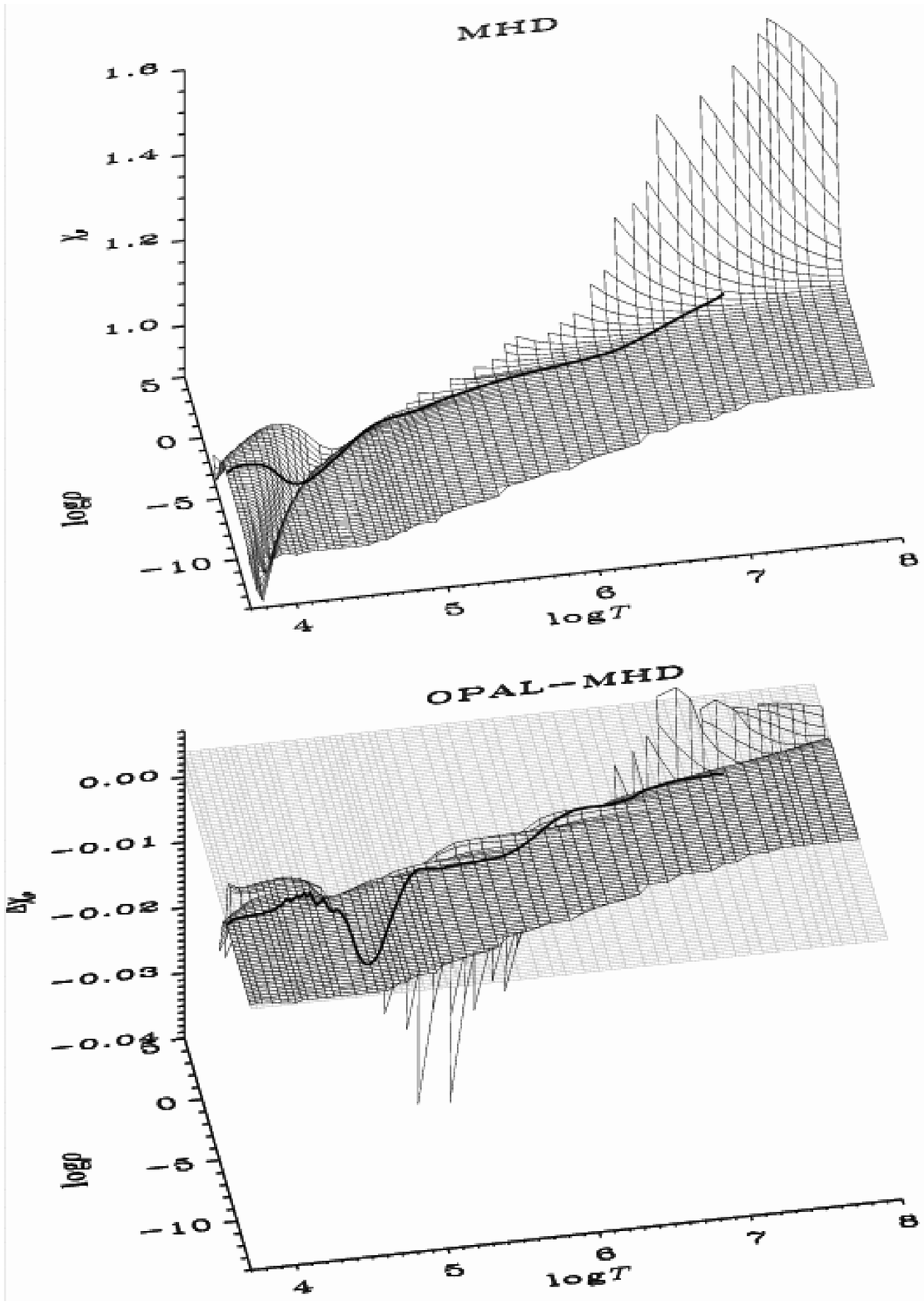}
\caption{$\chi_\varrho$, the logarithmic pressure derivative at constant
	temperature, for the full mixture in the upper panel, and its differences
	(OPAL minus MHD) in the lower panel.\label{table_H-Ne_norad3}}
\end{figure}

\begin{figure}
\plotone{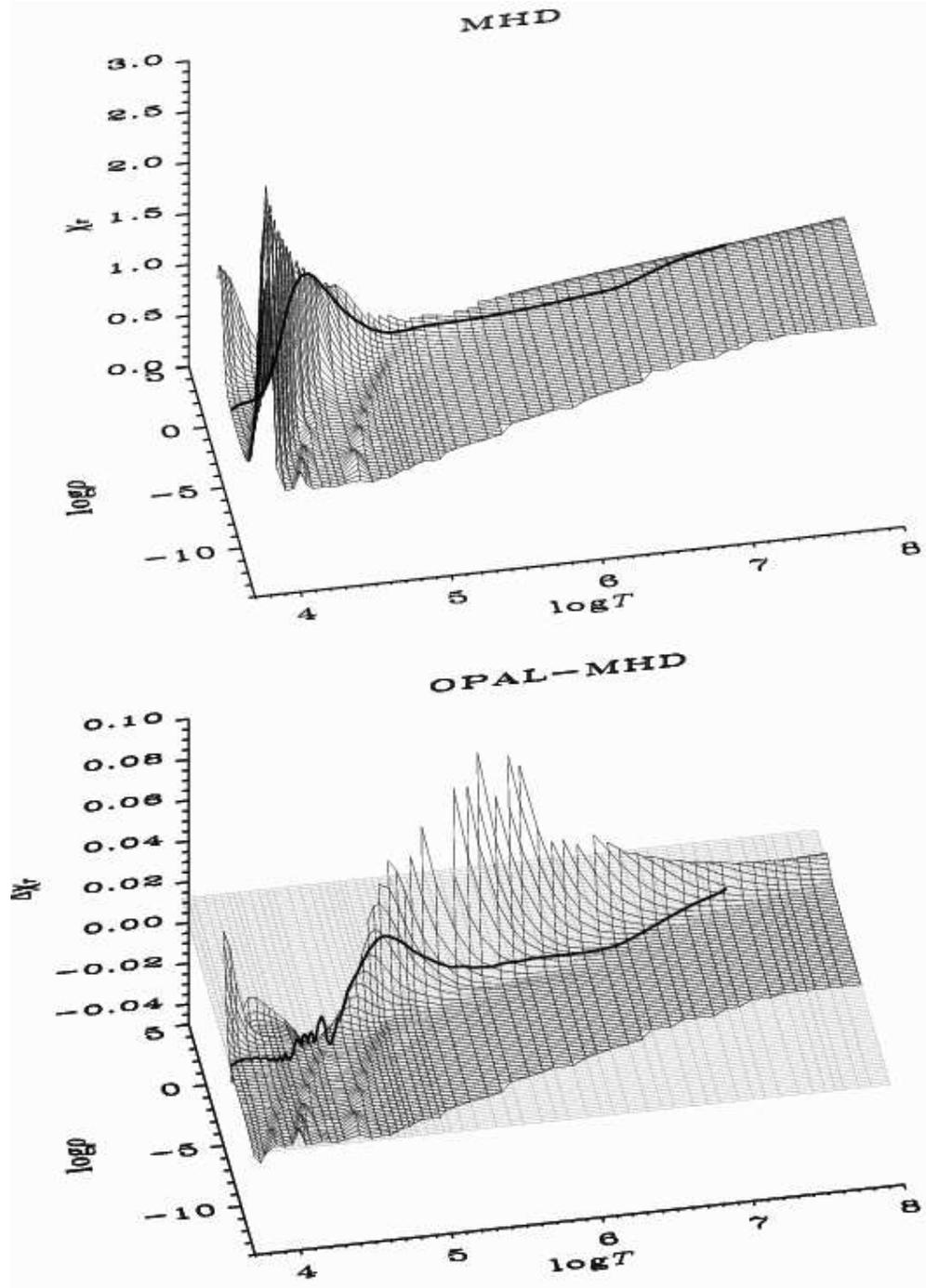}
\caption{$\chi_T$, the logarithmic pressure derivative with respect to
	temperature, for the full mixture in the upper panel, and its differences
	(OPAL minus MHD) in the lower panel.\label{table_H-Ne_norad4}}
\end{figure}

In this section we add the last four elements considered, namely 
carbon, nitrogen,
oxygen and neon. 
Comparing Figs.\ \ref{table_H-Ne_norad1}-\ref{table_H-Ne_norad4}
of this section with the corresponding
Figs.\ \ref{table_HHe_norad1}-\ref{table_HHe_norad4} of the previous section, 
hardly any differences appear,
neither in the absolute values nor in the differences between the two EOS.
\placefigure{table_H-Ne_norad1}

For a few points on the high-$R$ boundary of the tables,
differences in $\chi_\varrho$ and $\gamma_1$ have increased dramatically.
At least some of these odd points are the same for $\chi_\varrho$ and
$\gamma_1$. This might indicate that these points are spurious, possibly
associated with convergence problems in either EOS in this difficult region.
\placefigure{table_H-Ne_norad3}

\begin{figure}
\plotone{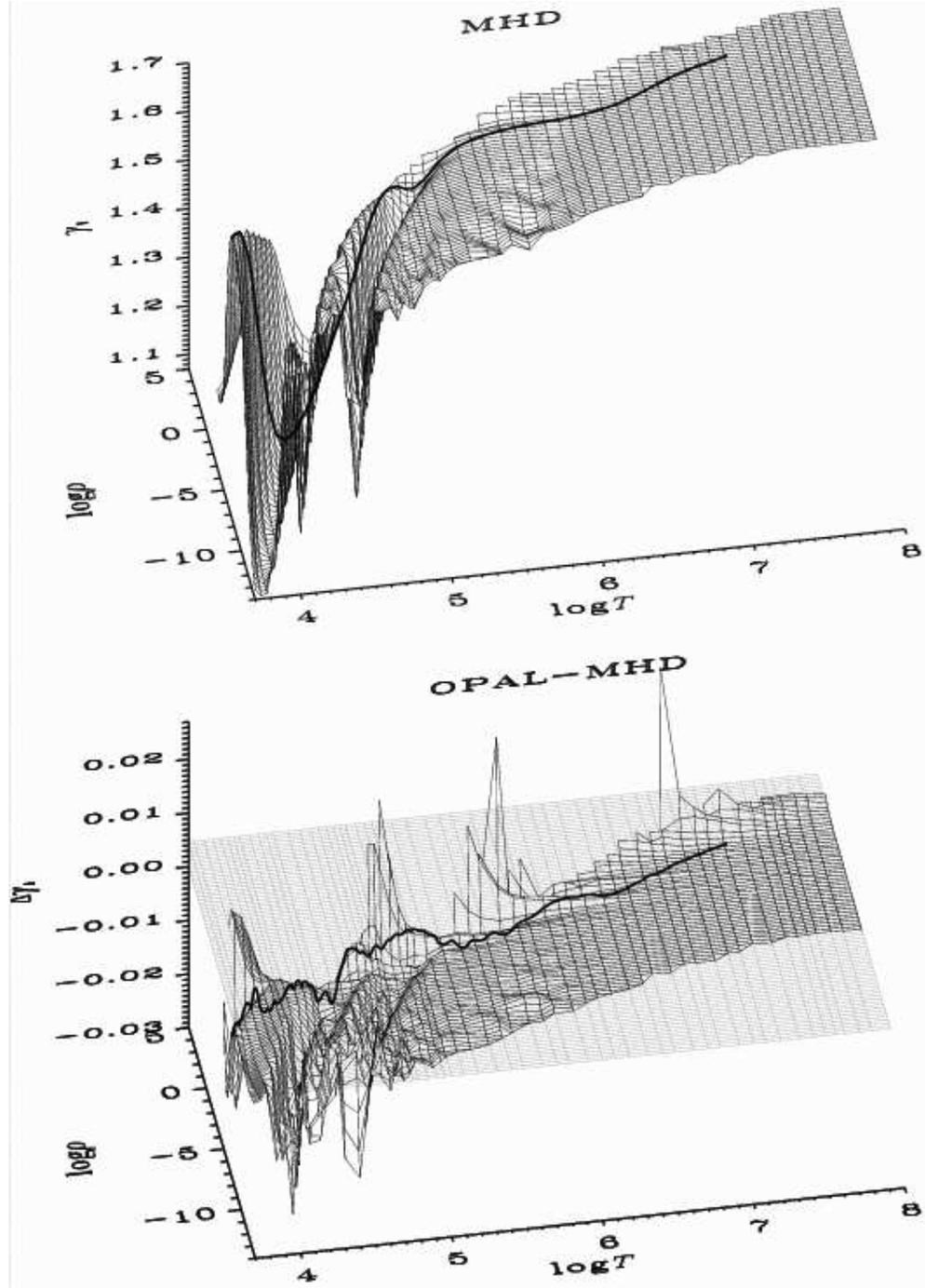}
\caption{The adiabatic logarithmic pressure derivative, $\gamma_1$, for the six
	element mixture in the upper panel, and its differences
	(OPAL minus MHD) in the lower panel.\label{table_H-Ne_norad8}}
\end{figure}

The heavy elements are just barely
discernible in the differences of $\chi_\varrho$
(Fig.\ \ref{table_H-Ne_norad4}). However, for $\gamma_1$, 
in Fig.\ \ref{table_H-Ne_norad8}, the
heavy elements appear clearly both in the absolute $\gamma_1$ and in the
$\gamma_1$ differences, especially along the low-$\varrho$ edge of the table.
Between the first and second ionization zones of helium ({\cf}
Fig.\ \ref{table_HHe_norad8}), we notice some wiggles, which are likely 
resulting from the third ionization
zones of carbon and nitrogen, and the second ionization zones of oxygen and
neon.
Above the second ionization zone of helium, we can see all the ionization 
zones from the fourth ionization zone of carbon right up to 
the tenth ionization of neon,
although they are not all resolved in this $\varrho$-$T$-grid.
A rough estimate reveals that the relative difference between MHD and OPAL for
the heavy elements is of about the same magnitude as the one
for the helium ionization
zones, {\ie} 3\%, or about 20 times worse than the 0.16\% agreement
for hydrogen.

This unexpectedly large discrepancy for the heavy elements
might be a hint that these differences are primarily caused by differences
in the lower excited states. For hydrogenic ions, there are 
analytical solutions for all
states. This might explain the small discrepancy for hydrogen. For ions with
more than one electron there are no analytical treatments, except 
for their higher
states, which become nearly hydrogenic. 
So it might well be that the lower lying states of the
non-hydrogenic ions are responsible for the differences noticed here. 
The Yukawa potentials \citep{yuk-pot-rf}, which are used to describe
bound electron states in OPAL, are fitted to give the correct (experimental)
ionization energies. MHD uses experimental results for
the energy levels. It is no surprise therefore to 
get quasi-perfect agreement on the
location of the ionization zones (confirmed by the
ridge-in-the-middle-of-the-valley picture in the $\gamma_1$ differences),
whereas the energies of lower lying
excitation levels might differ
These
differences propagate into
the partition functions
and affect the course of ionization. In addition,
\placefigure{table_H-Ne_norad4}
the differences in the adopted micro-field
distribution, and the mechanism by which they ionize highly excited states,
might play a r{\^o}le in this region \citep{nayfonov:EOS-high-n,Qmhd}.
Since the differences occur at the low-$\varrho$ edge of the table, we
expect however, that they mainly reveal differences in the thermal ionization,
not in pressure ionization.

Let us return to pressure and have a closer look at the non-ideal effects
in the high-$T$-high-$\varrho$ corner. From the dotted iso-$R$ lines in
Fig.\ \ref{table_H-Ne_norad1}, it is clear that the non-ideal effects 
are not functions of $R$
alone. Instead it turns out that they are largely functions of $\varrho^2/T^3$.
Comparing the perfect gas
pressure and the fully degenerate, non-relativistic electron pressure
\placefigure{table_H-Ne_norad8}
\begin{equation}
\nonumber
P_{\rm perf}       = \frac{\varrho \kB T}{\mu m_{\rm u}}\quad {\rm and}\quad
P_{\rm e}^{\rm deg} = \frac{1}{5}\left(\frac{3}{8\pi}\right)^{2/3}
			\frac{h^2}{m_{\rm e}}
			\left(\frac{\varrho}{\mu_{\rm e}m_{\rm u}}\right)^{5/3},
\end{equation}
we see that the two pressures
compete along $\varrho^2\propto T^3$-lines.
This means that relative to
high-$R$ (Coulomb) effects, there are more degeneracy effects 
in the high-$T$-$\varrho$ corner of the
table, which reveals 
the nature of the sharp rise of both $P$ and $\chi_\varrho$
in this region. The correlation with larger OPAL$-$MHD differences
(See lower panel of Fig.\ \ref{table_H-Ne_norad1})
prompted us to perform a direct comparison
between the Fermi-Dirac integrals from the two codes. We found non-systematic
differences a reassuring eight orders of magnitude smaller than the EOS
differences we observe in this region.

An alternative explanation could be the lack of electron exchange effects
in the MHD EOS. This is a combined effect of Heisenberg's uncertainty relation
\citep{heisenberg:uncert} and Pauli's exclusion principle \citep{pauli:excl}:
Due to the former, electron wavefunctions are
extended, but due to the latter, the wavefunctions of two close electrons with
same spin cannot overlap. This results in the combined wavefunction either
having a bulge or a node at the mid-point between the two electrons, giving
rise to two different kinds of contributions to the Coulomb interactions.
In the fully ionized, weak degeneracy limit, the first-order e-e exchange
pressure \citep{dewitt:thermodyn-deg-gas,dewitt:statmech-dense-gas} is negative and proportional to $\varrho^2/T$.
Analyzing the differences in solar solar case, we actually find in
Sect.\ \ref{EOSsun}, that those powers of $\varrho$ and $T$ are the ones best
describing the differences above $T\simeq 2\times 10^6\,$K
\section{Comparisons in the Sun}
\label{EOSsun}
To study the EOS under solar conditions, we have evaluated the MHD and OPAL EOS
on a $\varrho-T$ track that corresponds to the Sun using
the respective interpolation routines.
Obviously, this is a simplified comparison, not
of evolutionary models of the Sun, but merely of the equations of
state for fixed solar-like circumstances. 
As demonstrated elsewhere, such a simplified procedure is
well justified \citep[See {\eg}][]{jcd-Nature:osc-eos}.

We use the three chemical mixtures from
Table\ \ref{mixes}, bearing in mind that Mix~3 has about twice solar
metallicity.
In contrast to the comparisons of the previous section, we now
include radiative contributions.
This will of course not change the differences of thermodynamic
quantities, since both formalism use the well-known additive 
radiative contributions \citep{cox-guili:stel-struc1}.

In all the figures of this section, we notice that the MHD and the OPAL
EOS differs very little at temperatures below 20\,000\,K and above $10^6$\,K,
but they differ significantly in between. And though the differences are small
above 20\,000\,K, they are still of the same magnitude as the relativistic
effects which are well within reach of helioseismology \citep{elliott:RelDegSun}
---and they look intriguingly systematic.

The wiggles in the differences, most noticeable 
in the region between log$T$=4-4.5, 
are almost certainly due to the interpolation schemes. They
become quite dominant in the
$\gamma_1$ difference. As mentioned in Sect.\ \ref{EOSscape}, the tabular
resolution
of the OPAL tables for Mix~2 and 3 is about three times better than that of the
corresponding MHD tables, whereas the pure hydrogen tables (Mix~1) have the 
same (low) resolution.
Comparing the figures based on low resolution tables
(figs.~\ref{solar_H1}--\ref{solar_H8}) with those based on high resolution
table (Figs.~\ref{solar_HHe1}--\ref{solar_HHe8}), we notice that the
high-frequency component in the difference plots are much larger in the
former than in the latter. This is especially visible around
$\log T\simeq 4--4.5$ and $\log T\simeq 5.1--5.9$. This indicates that the OPAL
results have larger interpolation errors, since the MHD tables all have the
same resolution. A more detailed study of interpolation errors in tables,
was written by \citet{bahcall:dCZ}, for the case of OPAL {\em opacities}.
\subsection{Pure hydrogen}
\label{SunH}
If we take a look at the absolute pressure in Fig.\ \ref{solar_H1} a), we notice
\placefigure{solar_H1}
a bend at log$T=6.4$. This marks the bottom of the convection zone. Inside the
convection zone, that is below log$T=6.4$, there is
adiabatic stratification of pressure and temperature,
{\ie}
\begin{equation}
	\nabla_{\rm ad} = \dxdycz{\log T}{\log P}{\rm ad} =
	\frac{\gamma_1-\chi_\varrho}{\gamma_1\chi_T}.
	\label{nabla_ad}
\end{equation}
When the gas is nearly fully ionized, essentially 
$\nabla_{\rm ad} = 2/5$, evidenced as
the straight-line part of the curve in Fig.\ \ref{solar_H1} a).
In the ionization zones (the outer $0.02 R_\odot$),
$\nabla_{\rm ad}$ is lowered to about 0.11 at log$T\simeq 4.1$, again
clearly evidenced as
a depression in the pressure curve. $\nabla_{\rm ad}$ comes back to
$2/5$ at log$T\simeq 3.76$, but this happens at 
the top of the convection
zone where there is a downward bend to a radiative stratification.
\placefigure{solar_H3}
The {\em depth} of the convection zone is about 0.285$R_\odot$, and just
slightly higher, at a depth of 0.25$R_\odot$, hydrogen finally gets
fully ionized according to MHD (fewer than 1 in 10$^5$ are still neutral),
at a temperature
of log$T=6.3$. For comparison, MHD predicts that
hydrogen is 99.88\% ionized in the 
middle of the convection zone at log$T=6$. Both results are largely due to
pressure ionization, without which about a quarter of the hydrogen would still
be neutral at the solar center.
So, although
it is reasonable to say that the hydrogen ionization zone is confined to the
outermost 2\% of the Sun, one should also bear in mind the long 
tail of unionized hydrogen that is 
extending almost to the bottom of the convection zone. This tail
has an especially large effect on the opacity, since in the visual and UV
only bound states can
add opacity to the constant ``background opacity'' from
electron scattering.
\placefigure{solar_H4}

In the upper plot of Fig.\ \ref{solar_H3} we can actually see the differences
between the absolute values of $\chi_\varrho$. It is evident that
OPAL has a much
smoother and broader ionization zone than the somewhat bumpy MHD. Turning off
the $\tau$-correction (dashed lines), almost centers MHD on OPAL, but the bumps
remain the same. These bumps were also noticed by \citet{nayfonov:EOS-high-n}
and their analysis showed that in the region where log$T=4.2$--$5.2$ the bumps
are caused by excited states in hydrogen. In this part of the Sun, hydrogen
is 30\% increasing to 97.8\% ionized, so even a small amount of neutral
hydrogen can have a significant effect on the EOS.

At log$T\gtrsim 6.5$ we see how degeneracy
sets in, with the electron component of $\chi_\varrho$ increasing towards its
fully degenerate value
of $5/3$. In the lower plot, we notice that degeneracy is
accompanied by an increase in the differences. This could be 
attributed to the MHD EOS not including electron-electron exchange
effects, as pointed out in Sect.\ \ref{TabH-Ne}.
\placefigure{solar_H8}

The behavior of $\chi_T$ (Fig.\ \ref{solar_H4}) confirms the picture obtained
from Fig.\ \ref{solar_H3}, that is, MHD ionizing faster and 
being more bumpy than OPAL. However, since the
dynamic range of $\chi_T$ is much larger than that 
of $\chi_\varrho$, the bumps, which
have still about the same size as those of $\chi_\varrho$, are
now being dwarfed by the much larger ionization peak in $\chi_T$.
Comparing the differences shown in the lower plots, we notice that the
overall differences are about twice as large as for $\chi_\varrho$, but 
the ionization peak in the respective upper plots is about 10 times larger for
$\chi_T$ than for $\chi_\varrho$. 
We also notice that in $\chi_T$, 
as a likely result of the higher dynamic range, 
the interpolation-wiggles at log$T \lesssim 4.6$,
are much more prominent than in $\chi_\varrho$.

We can also distinguish MHD from OPAL in the absolute values of $\gamma_1$
(Fig.\ \ref{solar_H8} a)), although they are
much closer than in the $\chi$'s 
of Fig.\ \ref{solar_H3} and\ \ref{solar_H4}. This is
confirmed in the differences shown 
in panel b), which are overall smaller by an order of magnitude compared
to the $P$-, $\chi_\varrho$- and $\chi_T$-differences.
In contrast to our experience with $P$, $\chi_\varrho$ and $\chi_T$, 
here diminishing the $\tau$-correction in MHD
(dashed and dotted lines) does not lead to
any better agreement with OPAL. This is again convincing evidence that
$\gamma_1$ is a very efficient filter for high-$R$ effects.
The differences that we see are therefore due to the physics of ionization,
except at low temperatures, where interpolation errors seem to dominate.

\subsection{Hydrogen and helium mixture}
\label{SunHHe}
The effect of helium is very hard to see in the reduced pressure shown in 
Fig.\ \ref{solar_HHe1} a), and in the shape of the differences in
\placefigure{solar_HHe1}
Fig.\ \ref{solar_HHe1} b). However, a
comparison with the pure hydrogen case (Fig.\ \ref{solar_H1}), allows us 
none the less to see a few changes to the differences in the lower panels;
The peak around log$T=4.7$ gets considerable smaller by adding helium,
except for the $\tau=1$ case, where the difference actually increases in this
region. Also, the differences outside the high-$R$ region
decrease by adding helium, independently of the choice of $\tau$.

In general, adding helium does not alter the shape of the differences
in $P$, $\chi_\varrho$ or $\chi_T$, and the changes due to composition
are only manifest by
a change of the amplitude of the peak around log$T=4.7$.
This is surely due to the fact that most of the ionization in the Sun takes
place in the high-$R$ region, so that the first-order high-$R$
differences due to the ionizations themselves
\placefigure{solar_HHe3}
simply dwarf the second-order effects due to detailed partition functions,
among other. The solar track does follow the ionization
zones to some degree, and only enters the hydrogen ionization zone ``head on''.
With the solar track curving along the hydrogen ionization zone in this way
the ionization features will be smoothed out over a much larger temperature
interval than if we had examined an iso-chore. This smoothing leads to more
blending of ionization zones from various elements, hampering analysis.
The shape, merging and smoothing of the ionization features is best seen
in Figs.\ \ref{table_H-Ne_norad1}--\ref{table_H-Ne_norad4}.

This behavior is clearly illustrated in, {\eg}, 
$\chi_\varrho$ (Fig.\ \ref{solar_HHe3} a), 
where we observe a rather sharp onset of ionization followed by a
much slower transition to full ionization.
The second ionization of helium appears as part of
the bump around log$T=5$. The bump is somewhat more pronounced
than in the pure hydrogen case. A more careful comparison
\placefigure{solar_HHe4}
with the pure hydrogen case (Fig.\ \ref{solar_H3}) reveals the first
ionization zone of helium as a slight extension of the hydrogen peak, on
the side towards higher temperatures.
Helium gets almost fully ionized at log$T=6.0$, where 1.77\% is singly ionized
and 98.23\% doubly ionized. The ionization only proceeds gradually towards
higher temperature, until at
log$T=6.75$ it suddenly becomes fully ionized. This happens at a depth of
$0.63R_\odot$, at the edge of the hydrogen burning core.
At log$T=6.5$, just slightly above the temperature where hydrogen gets fully
ionized, there is finally no more neutral helium left.

For $\chi_T$ (Fig.\ \ref{solar_HHe4} panel a), the bump at
log$T\simeq 5$ is a clear sign of the helium added, as opposed to the
similar but more entangled bump in $\chi_\varrho$ ({\cf} Fig.\ \ref{solar_H3}
and \ref{solar_HHe3}).
The second He ionization zone is very distinct in $\gamma_1$
(Fig.\ \ref{solar_HHe8} a),
and the first ionization zone is manifested by 
a widening of the hydrogen ionization
zone towards the high-$T$ side. The differences (panel b) are just
as entangled as for pure hydrogen (Fig.\ \ref{solar_H8}) 
but with lower amplitude.
\placefigure{solar_HHe8}
On the descending part, just above log$T=5$, there are some
large interpolation errors, caused by the change to the coarser grid.
We also notice a peculiar bump at log$T=6.6$

Looking at the various difference plots in this section, we see a correlation
between a high amplitude in the differences and a high $R$-value, a property
we already inferred
from the solar track (Fig.\ \ref{table_H_norad1}). 
The minimum in $R$ is found at
the base of the convection zone, 
where we also find a local minimum in the magnitude of the differences
between the EOS. The location of this local minimum coincide for all four
thermodynamic quantities.
This confirms our suspicion that at least some of the discrepancy
stems from $\tau$.
The reason for this conjecture is that 
the differences between MHD EOS with different $\tau$ almost vanishes
in this region,
\placefigure{solar_H-Ne1}
whereas they increase in the same way as the MHD-OPAL differences grow for
intermediate temperatures.

At high temperatures, above a minimum occurring at log$T\simeq 6.4$,
the MHD-OPAL differences grow, but the differences between the three $\tau$
versions themselves remain small. 
On the solar track, log$R=-1.8$ at the minimum of the MHD-OPAL difference,
and it only rises slightly to -1.4 at log$T=6.8$ 
where the solar track bends to
follow more or less an iso-$R$ line. 
The constant $R$ value is attained around log$T=6.15$. 
The differences between the $\tau$-versions are indeed the same in both of
these regions (this is 
best seen in the pressure differences {\eg} Fig.\ \ref{solar_HHe1}), which
explains why the three curves with different $\tau$ follow each other 
so closely at high temperatures.
The MHD-OPAL difference in this region
\placefigure{solar_H-Ne3}
can therefore not be explained by the $\tau$-correction. It also turns out that
in this region, the differences of the $\varrho$-$T$ plane 
are mainly functions 
of $\varrho^2/T^3$ instead of $R$. In Sect.\ \ref{TabH-Ne}
we suggested that this dependence might 
arise from electron exchange effects or maybe from possibly different
evaluations of the Fermi-Dirac
integrals. However, a third explanation might be based on 
the quantum diffraction mentioned
in Sect.\ \ref{HigherOrder}.
\subsection{H,He,C,N,O and Ne mixture}
\label{SunH-Ne}
Adding C, N, O and Ne to the H-He mixture
has two main effects: first, it displaces 4\% He ({\cf}
Tab.\ \ref{mixes}), thereby diminishing the helium features, and second it
leads to a slight decrease in the high-$R$
OPAL-MHD differences due to the increased mean
charge [see Eq.\ (\ref{Lambda-approx})].
Only in $\gamma_1$ (Fig.\ \ref{solar_H-Ne8}), can
\placefigure{solar_H-Ne4}
the heavy elements be observed directly. Comparing with the H-He case
(Fig.\ \ref{solar_HHe8}),
and going from low to high temperatures, we first notice 
a slight diminishment of
the feature associated with the 
first ionization zone of helium due to the 4\% decrease of the
helium content.
This weakening of the He$^+$ feature
is counteracted by the second ionization zone of carbon (24.38\,eV), as well as
that of the less abundant Ne (21.56\,eV) and N$^+$ (29.60\,eV). 
The feature of the second ionization zone of helium (54.42\,eV)
is also diminished, but counteracted by the third ionization zone of oxygen
O$^{++}$ (54.93\,eV), the most abundant heavy element. 
C$^{++}$, C$^{3+}$, N$^{++}$ and Ne$^{++}$ adds further ionization in this
temperature region, but on either side of of the helium bump, effectively
smoothing the feature a little.

This feature is used for helioseismic determinations of
the Solar helium abundance \citep{basu-antia:Y}, as there are no helium lines
in the Solar (photospheric) spectrum. A change in the strength of this feature,
either from
a change in abundances (primarily oxygen) or a change in the equation of state,
will therefore have an effect on the Solar helium abundance. This is in
particular interesting as the proposed new Solar abundances by \citet{AGS05}
lowers $Z$ to about
\placefigure{solar_H-Ne8}
1.1\% and it will be discussed further in
Sect.~\ref{AGS05review}.

The helium signal in $\gamma_1$ is also shown as the solid curve in
Fig.~\ref{gam1_xcmp} where we have plotted the differences between Mix~2 and
Mix~3 with respect to Mix~1. The two helium ionizations zones are prominent
at $\log T\simeq 4.6$ and at $\log T\simeq 5$. There are also effects from the
20\% lower hydrogen abundance, especially at lower $T$. The dashed line
in Fig.~\ref{gam1_xcmp} shows the effect of adding the metals; reducing the
helium features, adding a broad feature around $\log T\simeq 6.2$ and
adding a ``continuum'' of features from there, and down to the second helium
ionization zone. Note that the helium features are not reduced in proportion
to the change in $Y$, indicating the metals that have ionization zones here.

Continuing in our exploration of $\gamma_1$ and going to
higher temperatures we notice a
slight straightening of the ``knee'' around log$T\simeq 5.3$, 
due to the intermediate
ionization stages of C, N and Ne with ionization
potentials between 47\,eV and 240\,eV. 
Finally, at log$T\simeq 6.2$, we find a broad dip supplied by
the two uppermost ionization stages of C,N,O and Ne, having 
ionization energies in the range between 400\,eV and 1\,360\,eV.

\placefigure{gam1_xcmp}
The only quantity in which the introduction of heavy elements is manifested 
directly is
$\gamma_1$, which is an important key variable in helioseismology (since it
is closely
related to adiabatic sound speed $c^2=\gamma_{1}p/\varrho$).
The promise of these features is that the presence of 
heavy elements is well marked in $\gamma_1$. Actually, this marking is
so distinct \citep{GDN:Z+partfnc-eff}, that in future solar and stellar
applications of the MHD and OPAL equations of state it might be
worth to include more heavy elements
The influence due to our small quantity of heavy elements 
is about three times larger than the
difference between OPAL and MHD, 
though we hasten to add that our heavy element abundance of $Z=4\%$
is chosen too high in order to exhibit the effects more clearly;
they would of course decrease with
a more solar metallicity around $Z=1$--$2$\%.

We have not discussed radiation pressure yet, merely
because of the lack of controversy about it. However, 
it is worth a few notes. The ratio between radiation
pressure and gas pressure is constant along iso-$R$ lines
the two being equal
around log$R\simeq -4.5$. The largest radiation effects therefore occur at
log$T=6.4$ where there is also the smallest discrepancy between OPAL and MHD.
The effect of radiation changes 
log$P$, $\chi_\varrho$, $\chi_T$ and $\gamma_1$ by
$0.0007, -0.001, 0.003, -0.002$, respectively.
\section{Discrepancies due to differentiation}
\label{diff}
A closer inspection of the derivatives in the perfect gas region reveals some
discrepancies which are likely 
due to the numerical differentiation performed in the OPAL EOS.
This is most noticeable in $\gamma_1$, where the OPAL-MHD
differences in the perfect gas region are as large as
0.03\%, which admittedly is small indeed.
Helioseismology will, however, soon be dealing with such precision.
This difference most probably comes
from problems in the numerical calculation of 
an adiabatic change as performed in OPAL (note that MHD uses 
\placefigure{gamma1}
essentially analytical expressions for $\gamma_1$,
$\chi_\varrho$ and $\chi_T$
Since an adiabatic change is not rectangular in the $T-\varrho$ plane,
such an interpretation is consistent with the fact that
the differences in the derivatives with respect to $\varrho$
and $T$ ($\chi_\varrho$ and $\chi_T$, respectively) 
are about one order of magnitude smaller.
This also means, that in the ionization zones where pressure and
entropy are non-linear functions of $\varrho$ and $T$, this
differentiation noise must be much larger.
On the other hand, the differences between OPAL and MHD are still
at least an order of magnitude larger than this differentiation noise.
We hope, however, that future improvements will make OPAL and MHD
converge to within that level of the actual EOS, requiring higher
numerical standards.

The differences in
$\chi_\varrho$ and $\chi_T$ have a
tendency to follow iso-$R$ tracks, while the differences in $\gamma_1$ follow
isotherms. These two behaviors are still unexplained.
In Fig.\ \ref{gamma1}, the differences following isotherms are pretty
clear, but the iso-$R$ differences are also visible, well below the rising
mountain at high $R$-values.
%
%
\section{New metal abundances}
\label{AGS05review}

Recently the previously converging Solar abundances \citep{GN92,GN96}, have
been upset by new abundance analysis by \citet{AGS05} (and references therein)
performed on 3D
radiation-coupled hydrodynamical simulations of convection in the Solar surface
layers \citep[See, {\eg},][]{stein:solar-granI}. This
approach completely avoids the concepts of micro- and macro-turbulence of 1D
models, as the line-broadening velocity-fields are included explicitly. This
results in synthetic spectral lines that not only have the widths of the
observed lines, but also match the detailed (asymmetric) shapes of the lines
\citep{asplund:solar-Fe-shapes}. This reduces the number of free parameters
to essentially the abundances we seek, and gives less ambiguous abundances.
The result is a general lowering of the Solar heavy element abundances \citep{AGS05}
and most notably a halving of the oxygen abundance. This has a severe impact
on the previously close agreement between Solar models and helioseismic
observations \citep[See, {\eg},][and references therein]{bahcall:seism-AGS05}.
The lowering $Z$ to about 1.1\%, decreases the opacity at the bottom of the
convection zone thereby decreasing the depth of the convection zone
\citep{bahcall:seism-AGS05,bahcall:newOP+AGS05} at odds with helioseismic
measurements \citep{jcd:dCZ,basu:dCZ,bahcall:dCZ}.

In Sect.~\ref{SunH-Ne} we touched upon the determination of the helium abundance
by helioseismically measuring the bump in $\gamma_1$ around $T=10^5\,K$. This
bump also has a contribution from oxygen, as shown in Fig.~\ref{gam1_xcmp},
and a lowering of the oxygen abundance
will therefore have an effect on the Solar helium abundance.
To keep the agreement with the helioseismic helium abundance determination,
a smaller $Z$ has to be accompanied by a an increase in $Y$. 
We have interpolated the $\gamma_1$-differences of Mix~2 and 3 with respect
to Mix~1, to find the amplitude of the feature around $\log T = 5$.
From this we find
that a change of $Z$ from 1.8\% to 1.1\% should be accompanied by an
increase of $Y$ by 0.0039 (using MHD) in order to keep the bump unchanged.
{\bf
\citet{basu:SolarZ} perform a helio-seismic determination of the Solar
helium abundance using the new and lower $Z$, but based on the updated
OPAL EOS \citep{rogers:newOPAL}. They use a slightly higher $Z=1.26$\% for
the new abundances and find an increase in $Y$ of 0.0050. To translate to
our choices of $Z$ we derive a $\Delta Y/\Delta Z = -1.029$ from their
results, which gives a $Y$-increase of 0.0072. This means that the $\gamma_1$
of the OPAL EOS is about 1.8 times more sensitive to the heavy element content,
than the MHD EOS is.
}
Evolutionary models with the two sets of abundances,
calibrated to the present Solar radius and luminosity, on the other hand, result
in a {\em decrease} of $Y$ by 0.005 \citep{bahcall:seism-AGS05}---a total
discrepancy between models and helioseismology of about 0.0089, accompanied
by large discrepancies in the depth of the convection zone and sound speed and
density profiles. For comparison, the helioseismic method for measuring the
amplitude of the He-bump, has an uncertainty of only 0.001-0.002.
It is worth mentioning here, that \citet{basu:Rsun-err} find
that inversions using the MHD equation of state results in $Y$ being about
0.004 higher than when using OPAL.
{\bf
This is most likely a consequence of our
finding above, that the heavy elements make a larger contribution to $\gamma_1$
in the OPAL EOS, resulting in a lower helium abundance compared to the
same analysis based on the MHD EOS.
}
The difference between OPAL and MHD in this region, although at a maximum here,
is therefore too small to solve the problem. This
is, however, also a region of the Sun with rather extreme plasma conditions,
as measured by the coupling parameter (See Eq.~[\ref{Gamma}]) making this a
likely site for further improvements to the equation of state.

\citet{lin:seismEOS} performed a very interesting analysis of response of the
intrinsic $\gamma_1$ to a range of abundance changes, and the ability of
helioseismology to pick-up these changes. The intrinsic $\gamma_1$ differences
between a model and the Sun, are the differences between $\gamma_1$s reduced
to the same temperature and density stratification, {\ie}, subtracting the
effects of different stratifications and finding the intrinsic differences
\citep{sb-jcd:eos-osc}, as
caused by equation of state differences or abundance differences, and therefore
directly comparable to our Figs.~\ref{solar_H8}, \ref{solar_HHe8} and \ref{solar_H-Ne8}.
\citet{lin:seismEOS} find that a change in $Y$ of 0.01 for fixed $X$, results in
changes in $\gamma_1$ of similar magnitude as we see between MHD and OPAL in
the present work. Furthermore, a decrease in the carbon abundance actually
improves the agreement between the Sun and models below $r=0.95R_\odot$. Both
effects are visible in helioseismic inversions of the models.
Bear in mind that these abundance changes are accompanied by changes to the
sound speed and density too, and we are therefore still far from a reconciliation
between the new abundance determinations and the helioseismic inversions.

An examination of the new opacities from the Opacity Project
\citep{seaton:OP-OPAL04,OP05}
shows that the sensitivity to changes in the helium abundance is minimal
in the region around the bottom of the convection zone. The change in helium
opacity is compensated by a similar, but opposite change in the hydrogen and
heavy element opacities, and the depth of the convection zone will therefore be
unaffected by a small increase of the helium content.
%
%
\section{Conclusion}
\label{conclusion}
\begin{figure}
\plotone{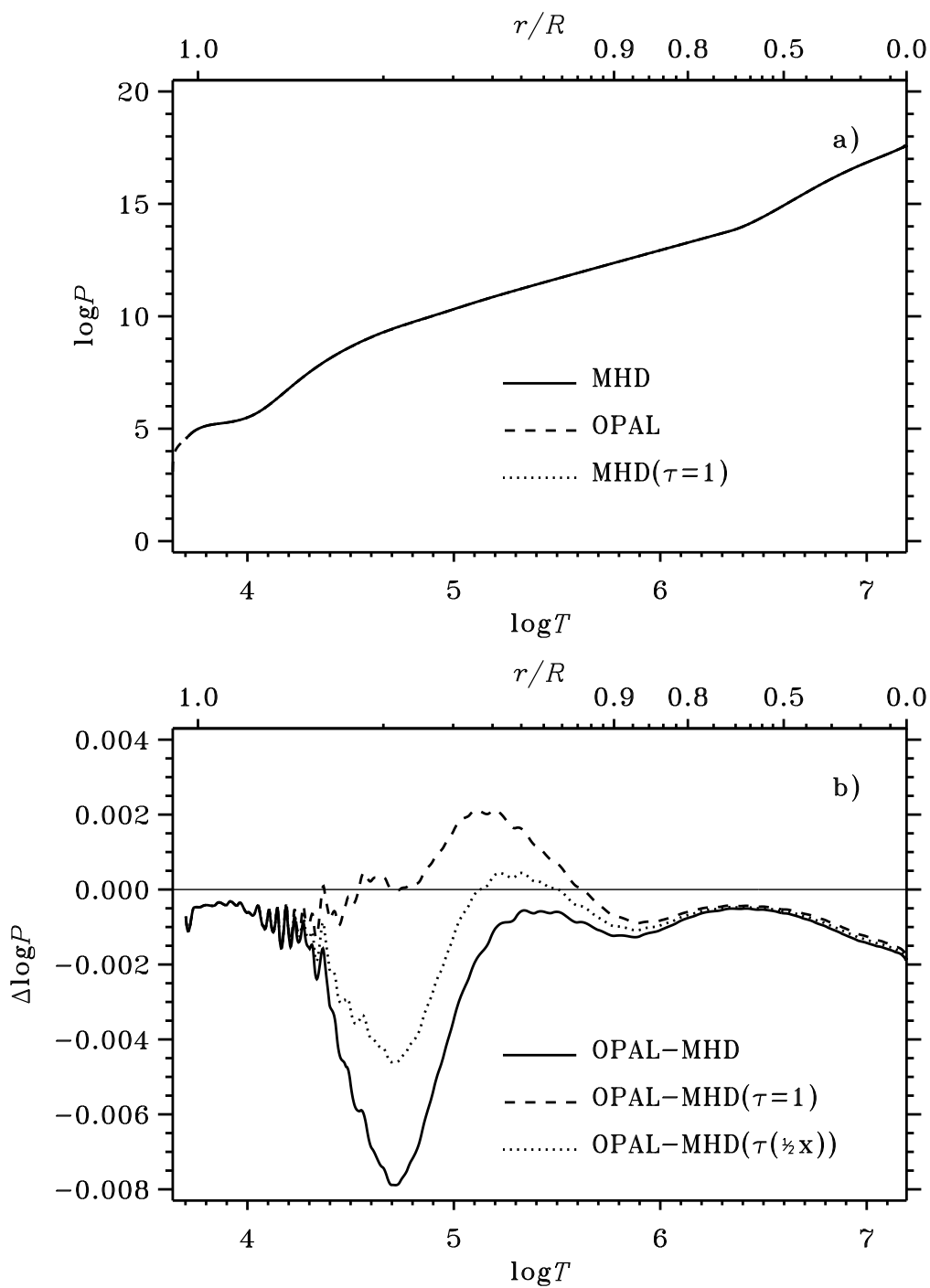}
\caption{The logarithmic pressure along a solar $\varrho,T$-track for pure
	hydrogen. The upper panel shows the absolute values of the
	MHD (solid line) and the OPAL (dashed line) pressure. We also plot the MHD
	pressure, using $\tau=1$ to show the effect of omitting this correction
	({\cf} Sect.\ \ref{tau}). These three pressures are indistinguishable unless
	we look at the lower plot, showing the difference OPAL minus MHD. Here we
	show, apart from the normal MHD, also the version with $\tau=1$, which seems
	closer to OPAL, and a version where we have halved the argument of $\tau$.
	\label{solar_H1}}
\end{figure}

\begin{figure}
\plotone{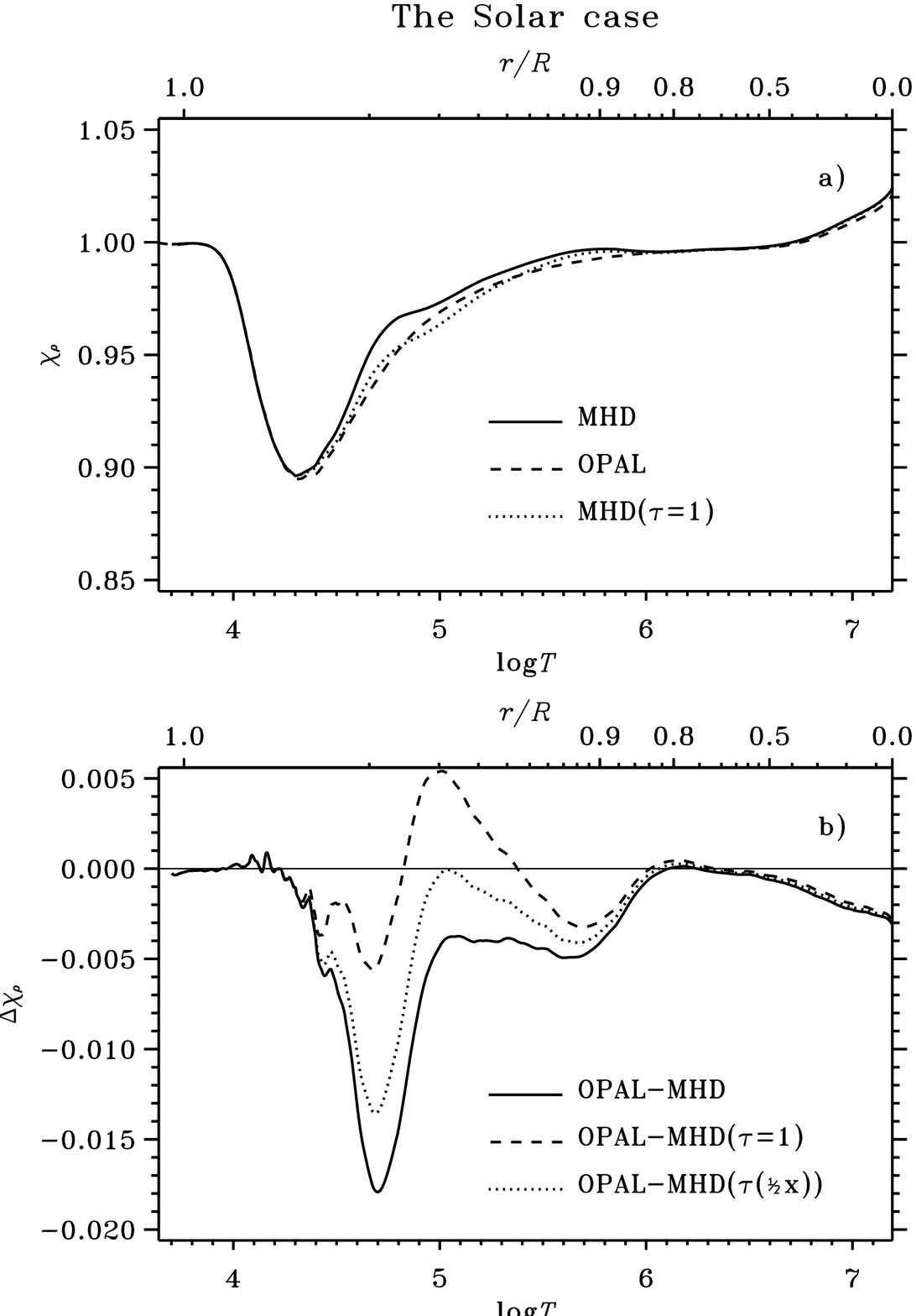}
\caption{The logarithmic pressure derivative with respect to density,
	$\chi_\varrho$, along the solar track for pure hydrogen.
	{\bf a)} the absolute value, {\bf b)} the difference (OPAL minus MHD).
	\label{solar_H3}}
\end{figure}

\begin{figure}
\plotone{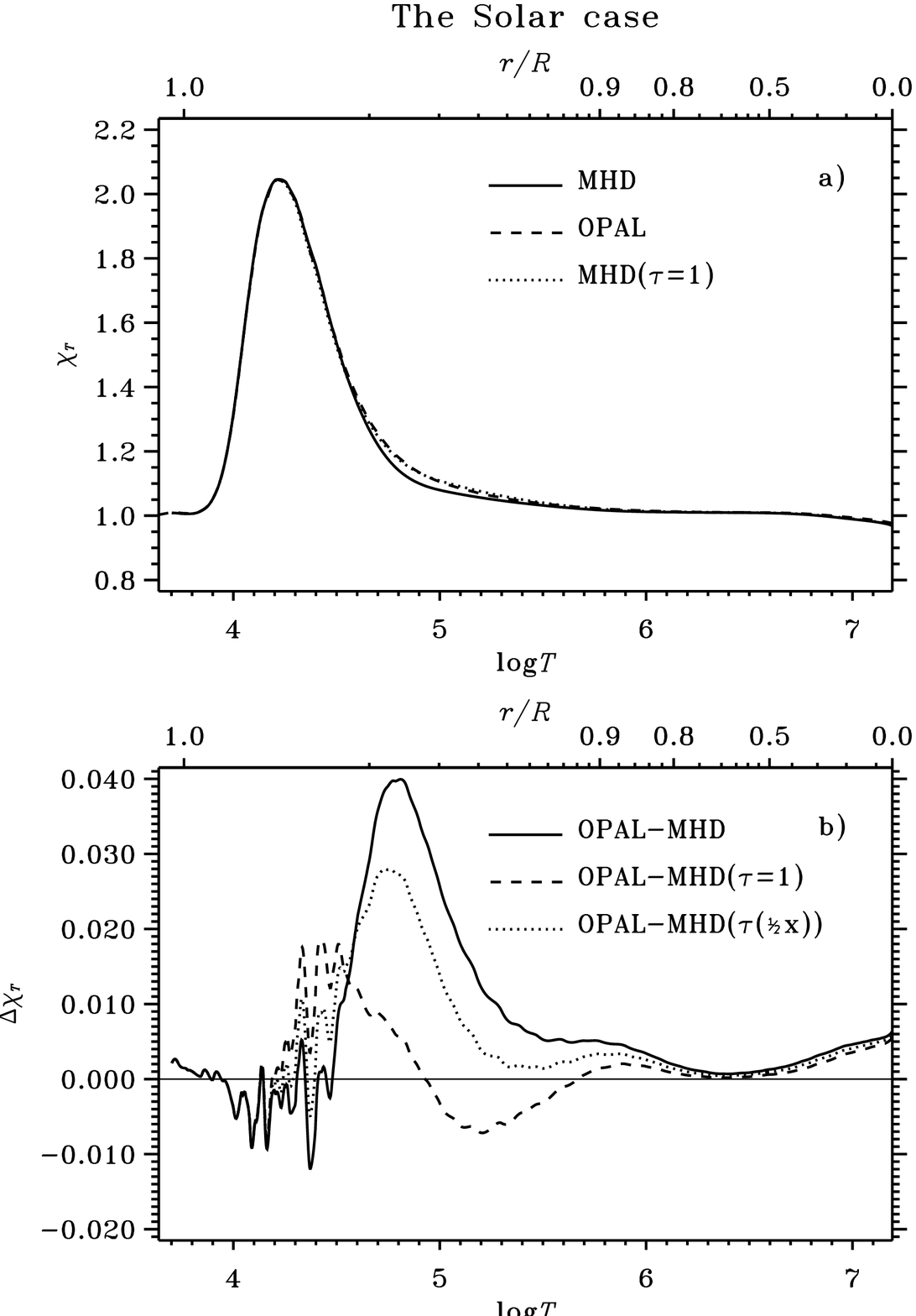}
\caption{The logarithmic pressure derivative with respect to temperature,
	$\chi_T$, along the solar track for pure hydrogen.
	{\bf a)} the absolute value, {\bf b)} the difference (OPAL minus MHD).
	\label{solar_H4}}
\end{figure}

\begin{figure}
\plotone{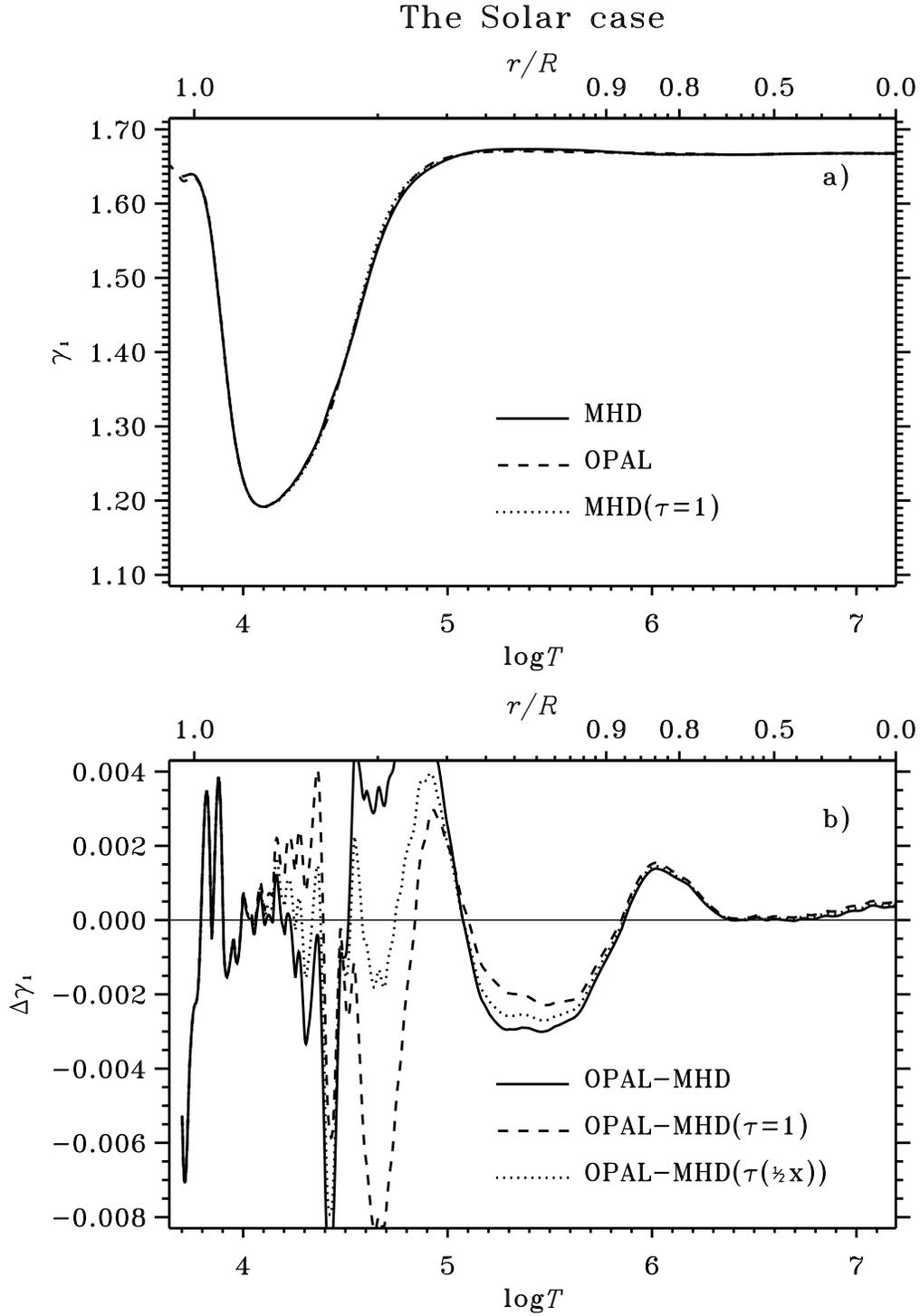}
\caption{The adiabatic logarithmic pressure derivative,
	$\gamma_1$, along the solar track for pure hydrogen.
	{\bf a)} the absolute value, {\bf b)} the difference (OPAL minus MHD).
	\label{solar_H8}}
\end{figure}

\begin{figure}
\plotone{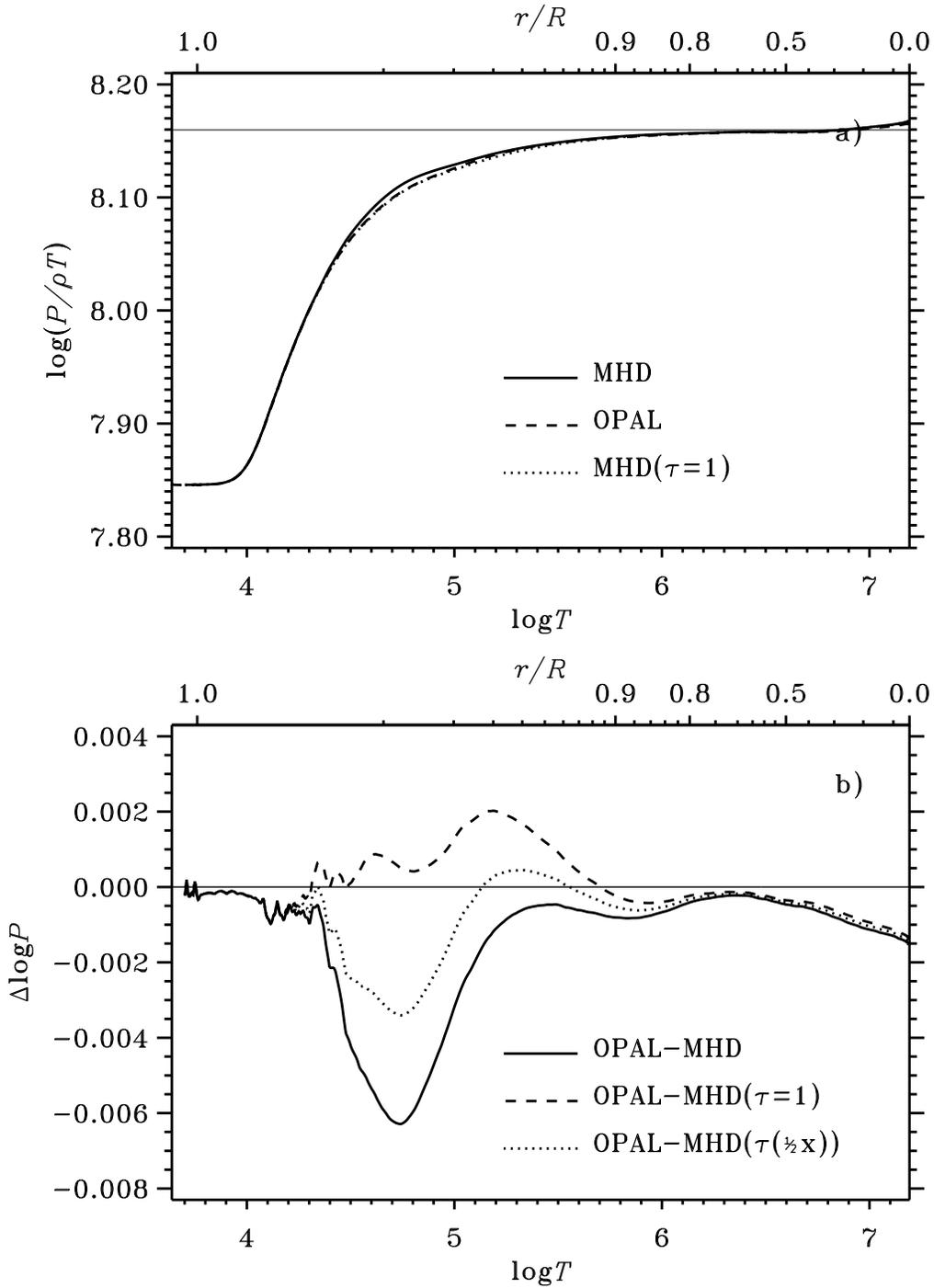}
\caption{The reduced pressure in the H-He-mixture along the solar track.
	{\bf a)} the absolute value, {\bf b)} the difference (OPAL minus MHD).
	The thin horizontal line in panel a), indicates the fully ionized, perfect
	gas pressure.\label{solar_HHe1}}
\end{figure}

\begin{figure}
\plotone{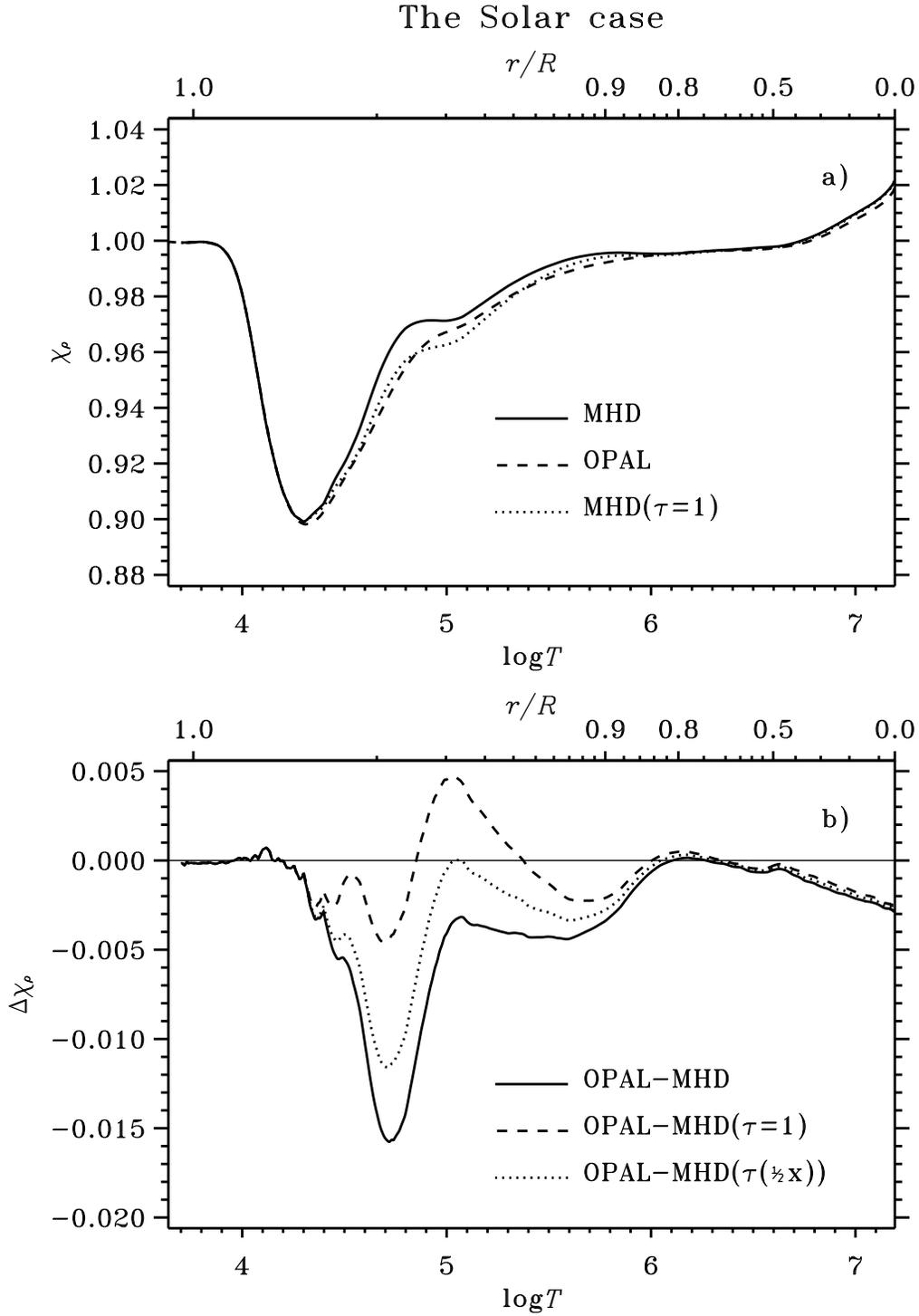}
\caption{$\chi_\varrho$ for the H-He-mixture along the solar track.
	{\bf a)} the absolute value, {\bf b)} the difference (OPAL minus MHD).
	\label{solar_HHe3}}
\end{figure}

\begin{figure}
\plotone{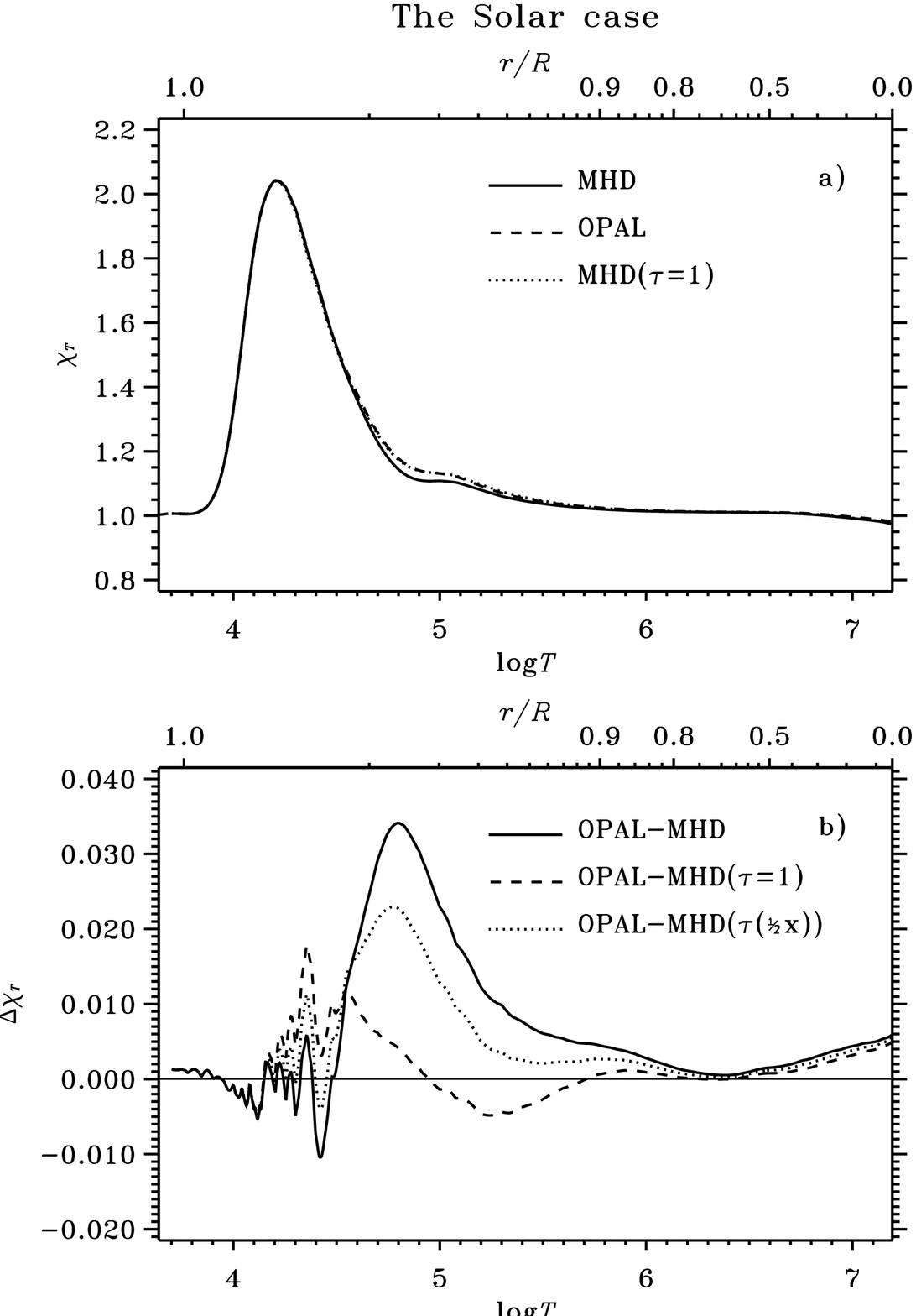}
\caption{$\chi_T$ for the solar track and the H-He-mixture.
	{\bf a)} the absolute value, {\bf b)} the difference (OPAL minus MHD).
	\label{solar_HHe4}}
\end{figure}

\begin{figure}
\plotone{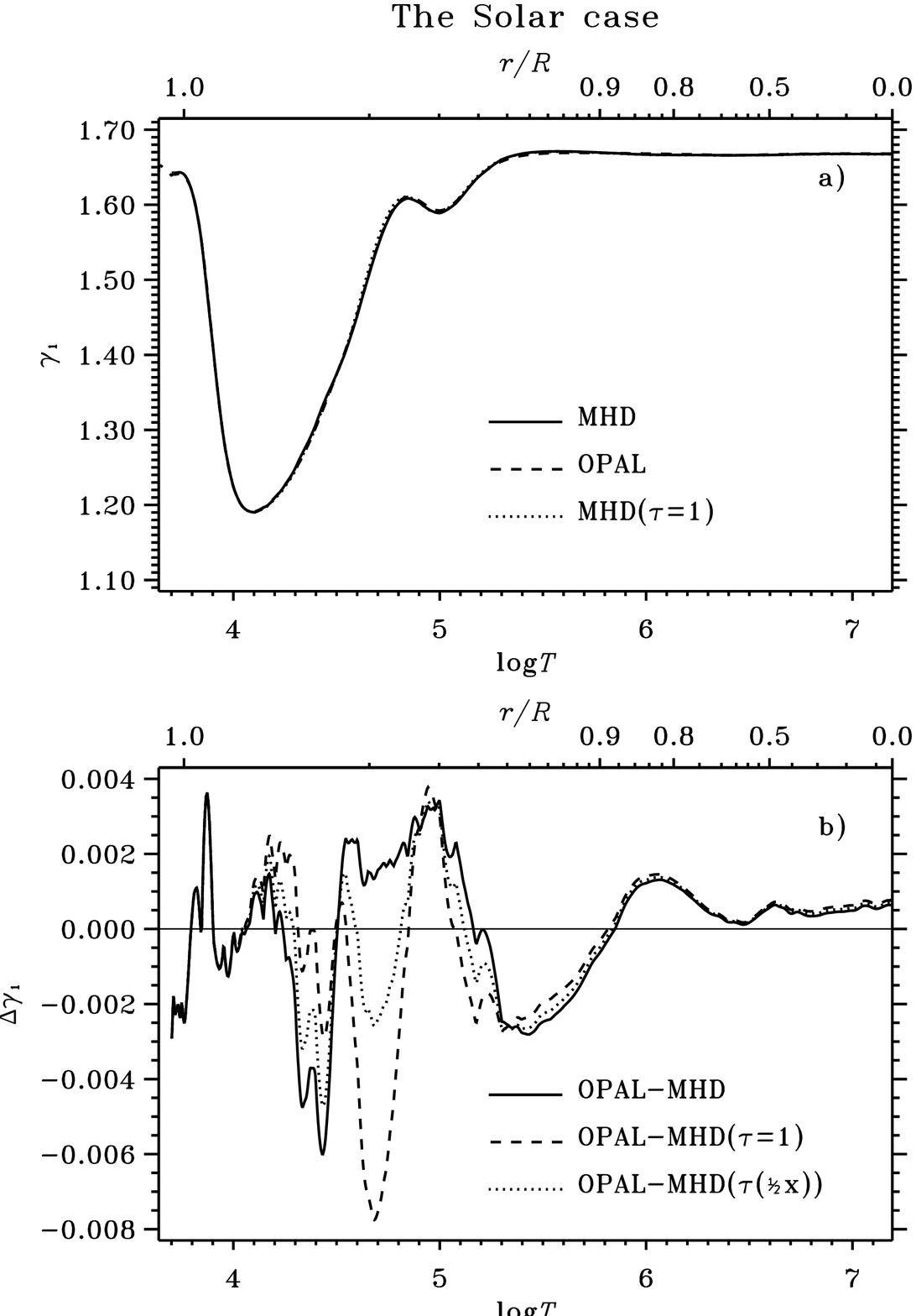}
\caption{$\gamma_1$ for the solar track and the H-He-mixture.
	{\bf a)} the absolute value, {\bf b)} the difference (OPAL minus MHD).
	\label{solar_HHe8}}
\end{figure}

\begin{figure}
\plotone{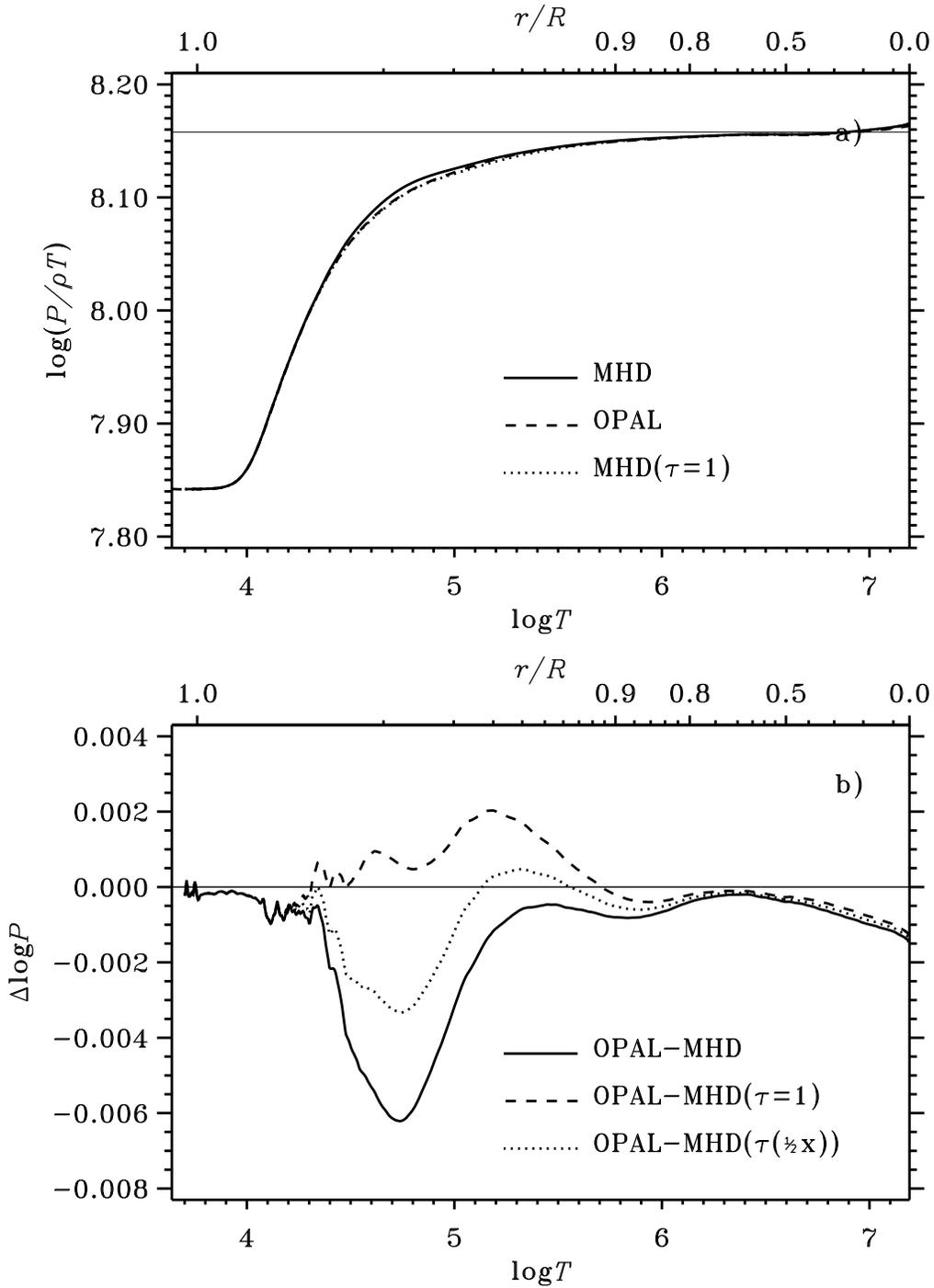}
\caption{Reduced pressure for the solar track and the 6-element mixture.
	{\bf a)} the absolute value, {\bf b)} the difference (OPAL minus MHD).
	The thin horizontal line in panel a), indicates the fully ionized, perfect
	gas pressure.\label{solar_H-Ne1}}
\end{figure}

\begin{figure}
\plotone{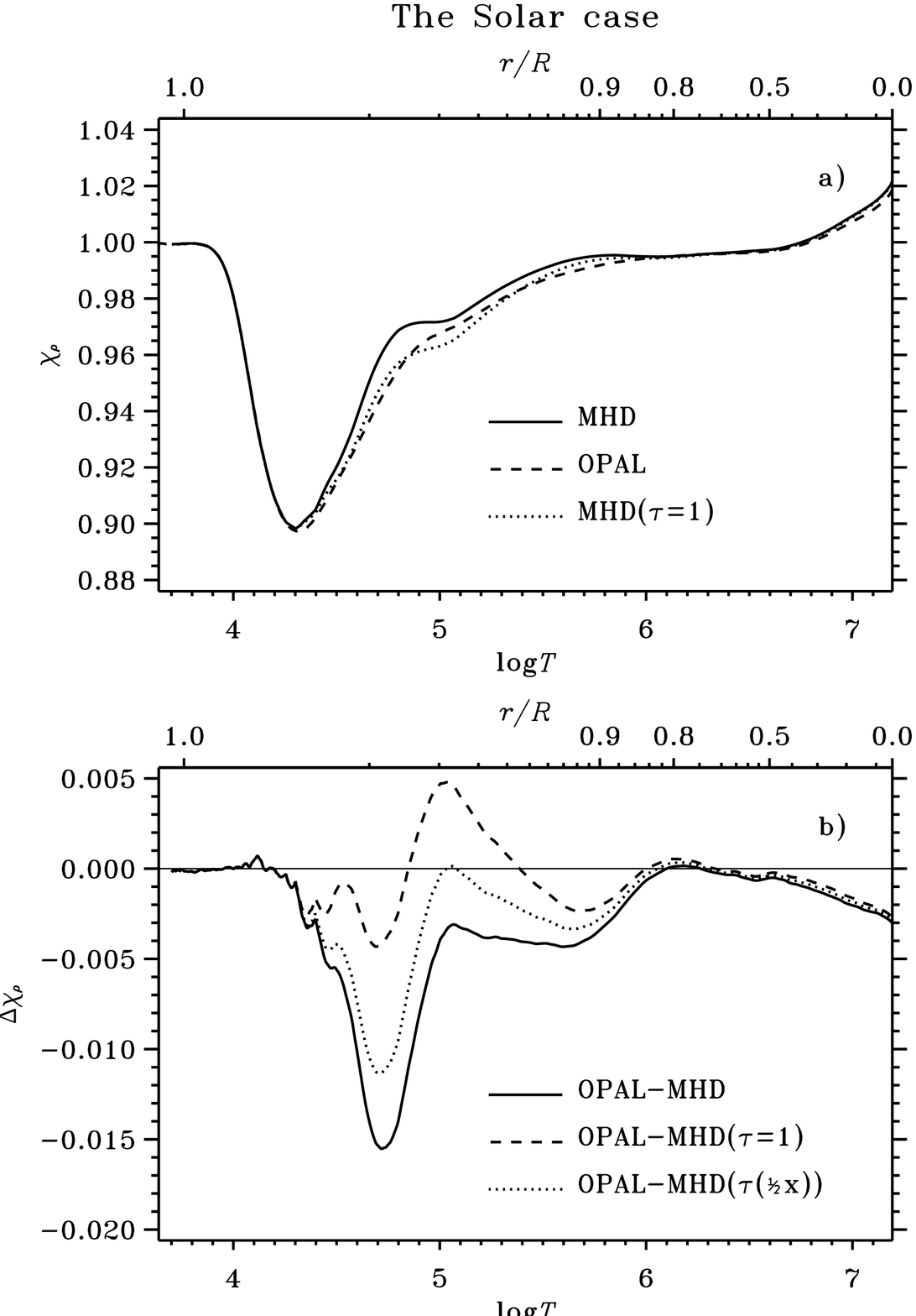}
\caption{The logarithmic pressure derivative with respect to density
	$\chi_\varrho$ for the 6-element mixture along the solar track.
	{\bf a)} the absolute value, {\bf b)} the difference (OPAL minus MHD).
	\label{solar_H-Ne3}}
\end{figure}

\begin{figure}
\plotone{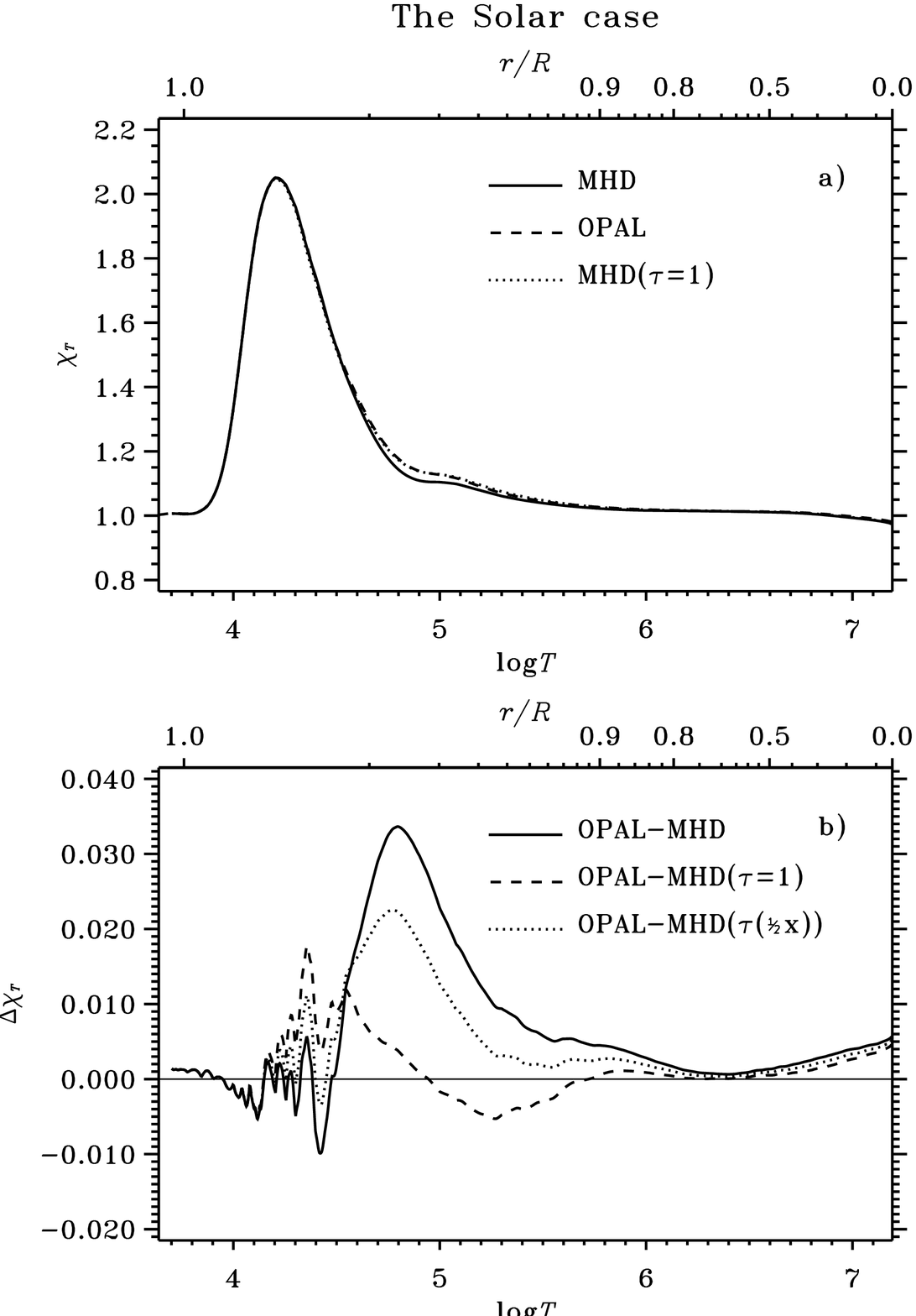}
\caption{The logarithmic pressure derivative with respect to temperature
	$\chi_T$ for the 6-element mixture along the solar track. {\bf a)} the
	absolute value, {\bf b)} the difference (OPAL minus MHD).
	\label{solar_H-Ne4}}
\end{figure}

\begin{figure}
\plotone{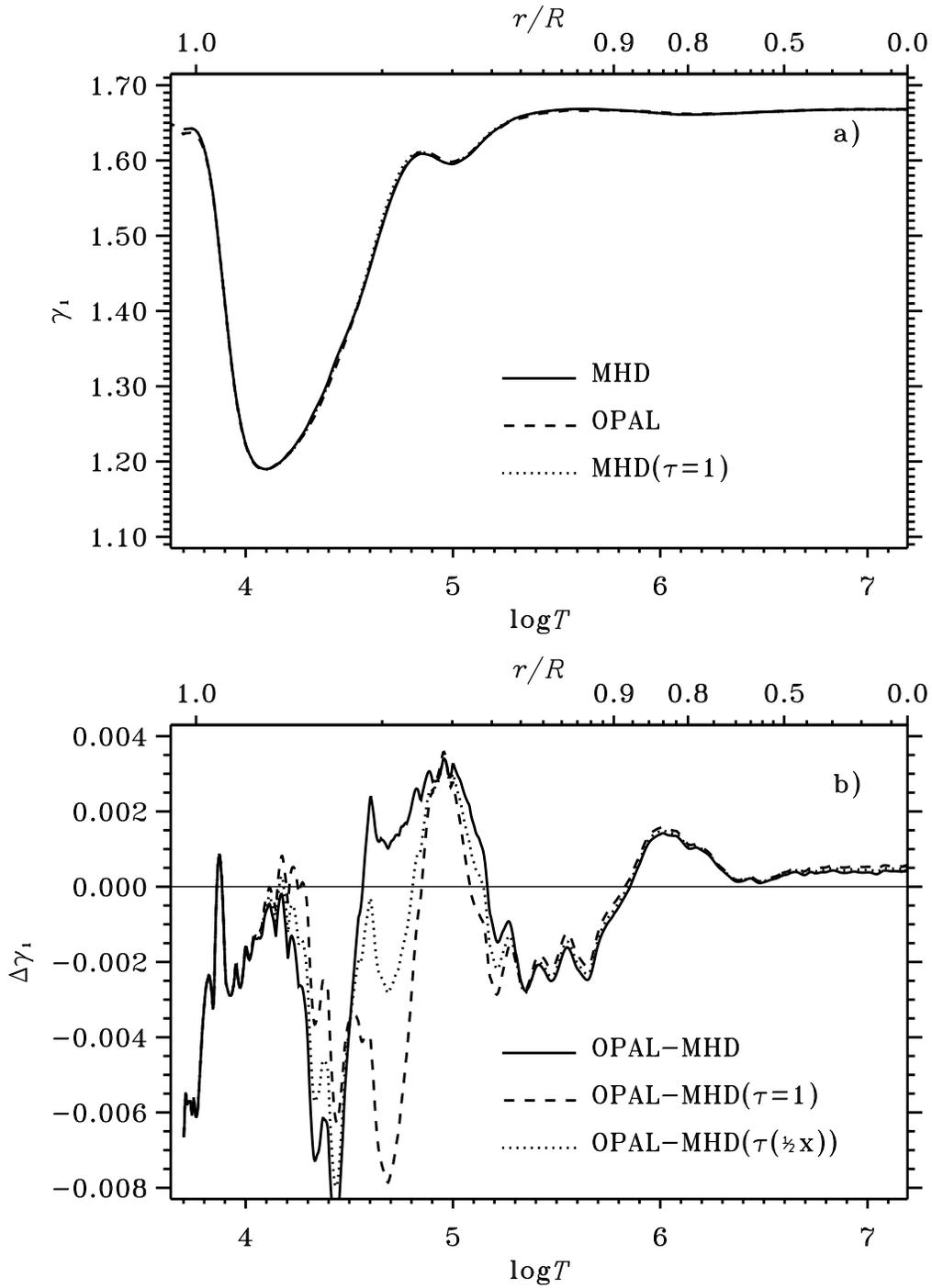}
\caption{The adiabatic logarithmic pressure derivative, $\gamma_1$, for
	the six-element
	mixture along the solar track. {\bf a)} the absolute value, {\bf b)} the
	difference (OPAL minus MHD).\label{solar_H-Ne8}}
\end{figure}

\begin{figure}
\plotone{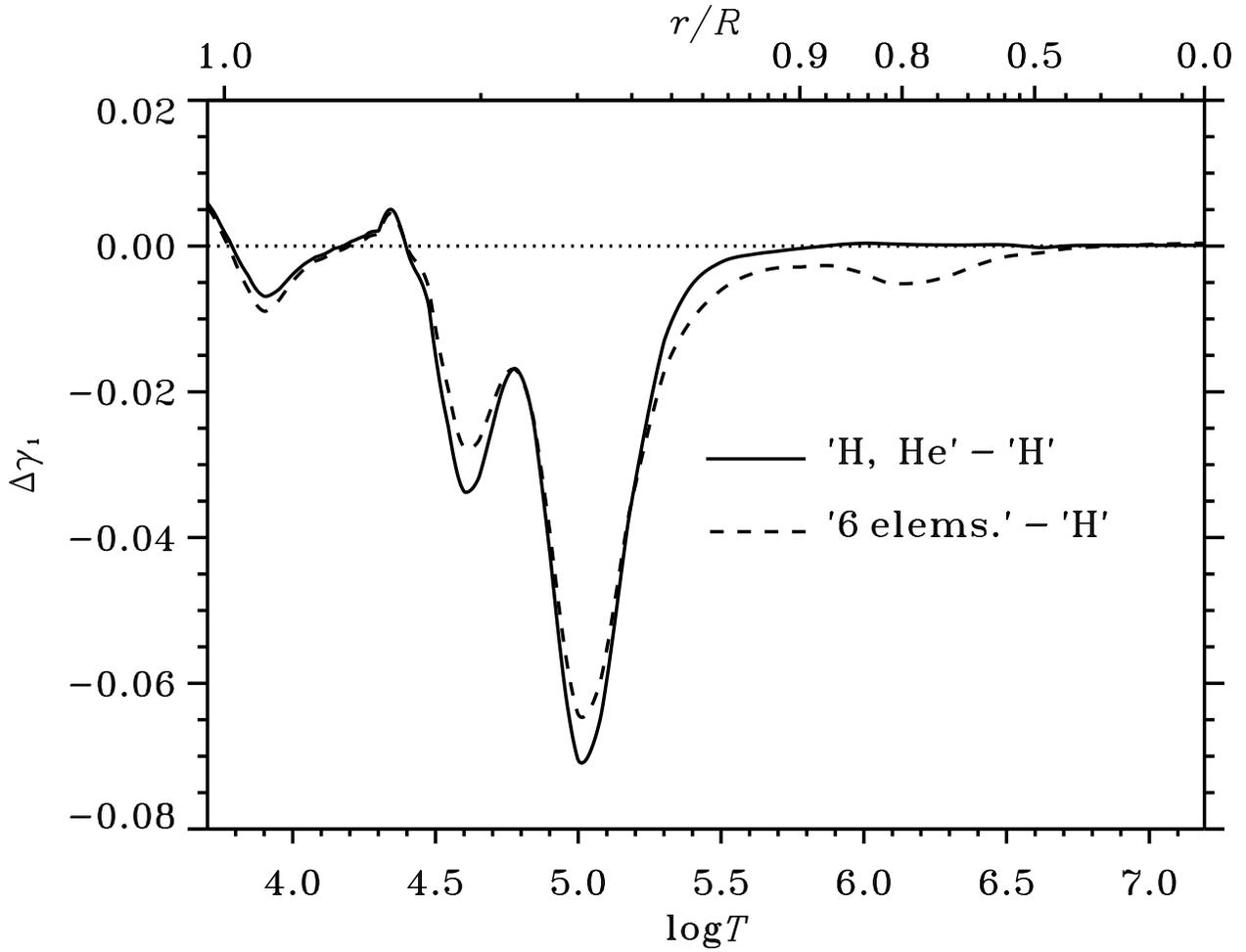}
\caption{The differences between $\gamma_1$ for the three different chemical
    mixtures along the solar track. The solid lines shows the Mix~2$-$Mix~1
	difference and the dashed line is the Mix~3$-$Mix~1 difference.
	\label{gam1_xcmp}}
\end{figure}

\begin{figure}[htb]
\plotone{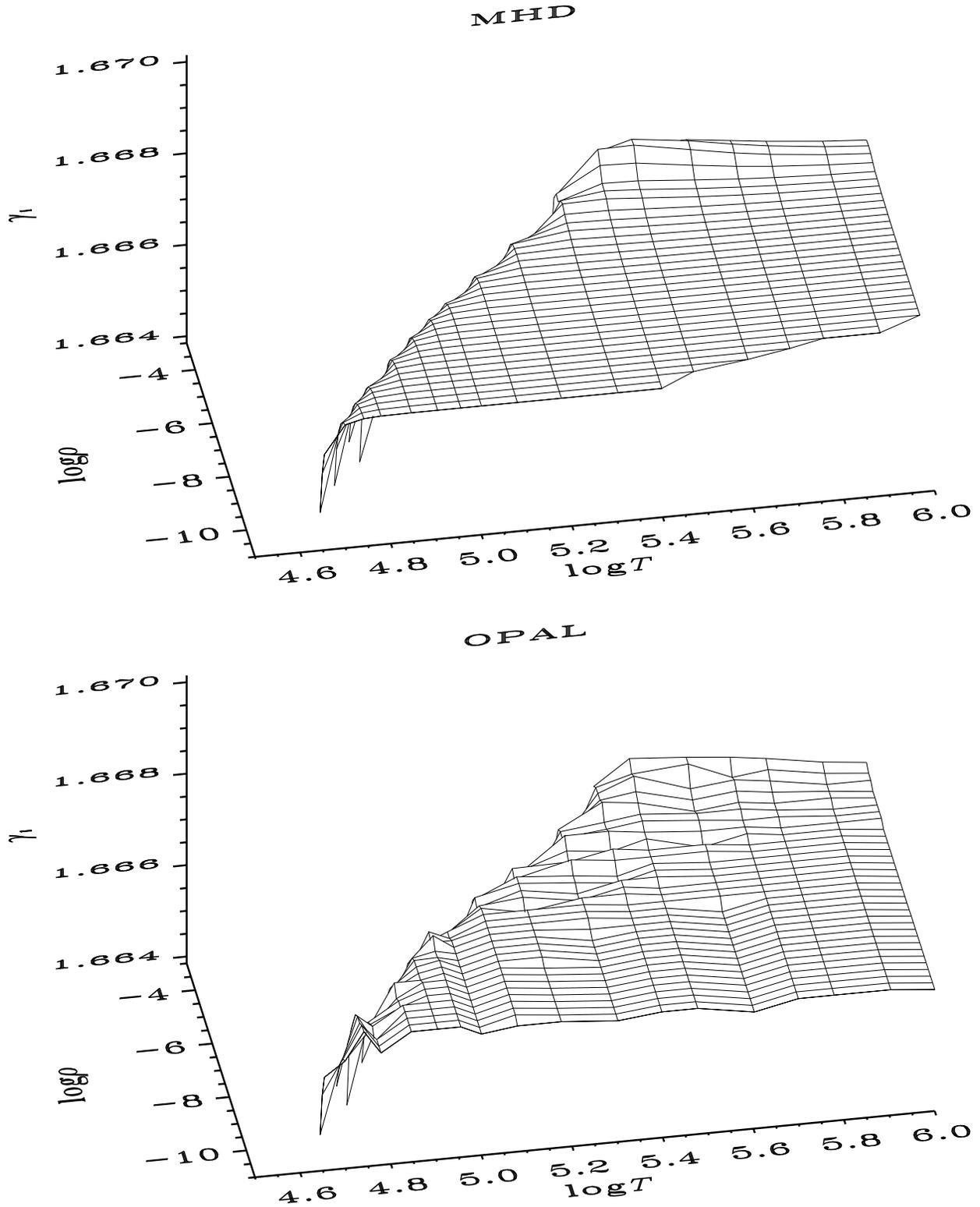}
\caption{This is a zoom-in on the fully ionized, perfect gas region of a
	pure hydrogen plasma ({\cf} Fig.\ \ref{table_H_norad8}), where
	$\gamma_1=5/3$. The upper panel shows the results for the MHD EOS
	which uses analytical expressions for all first- and second-order
	derivatives. The lower panel shows the same for the OPAL EOS, where
	derivatives are calculated numerically on a grid that are much denser in
	$\varrho$ and $T$ though, than in the tables published.\label{gamma1}}
\end{figure}

The present comparison of the two MHD and OPAL EOS has revealed
the reasons of several differences between these equations of state. 
They
can be summarized as follows (in order of importance):
\newcounter{nr}
\begin{list}{\alph{nr})}
  {\usecounter{nr}}
\item We find the largest differences at high densities and low temperatures,
	  or more precisely, at high $R$-values. From Sect.\ \ref{DHapprox} and
	  Eq.\ (\ref{Lambda-approx}) we know
	  that this property is 
	  indicative of differences in the treatment of plasma
	  interactions. Comparing the peaks of the differences in {\eg} pressure
	  (See Figs.\ \ref{solar_H1}, \ref{solar_HHe1} and\ \ref{solar_H-Ne1}), 
	  we obtain
	  Mix~1-to-Mix~2 ratios of 0.797, and Mix~1-to-Mix~3 ratios of 0.788, 
	  which
	  agrees very well with Eq.\ (\ref{Lambda-approx}), 
	  and thus further substantiates
	  our interpretation.
	  These differences are lowered
	  dramatically when we set $\tau=1$ in MHD, 
	  indicating that it is worthwhile to abolish $\tau$ and
	  reconsider how to get rid of the short-range divergence in the
	  plasma-potential (See Sect.\ \ref{HigherOrder}).
\item In the high-temperature-high-density corner of the tables we observe how
	  degeneracy sets in. Along with degeneracy, we also notice how some
	  specific differences are growing. This effect could be due to quantum
	  diffraction or exchange effects, both included in OPAL but not in MHD. 
	  Quantum diffraction is
	  the effect of the quantum mechanical smearing out of, primarily, the
	  electron due to its wave nature. The exchange effect is a
	  modification of the quantum diffraction arising from the anti-symmetric
	  nature of two-particle wavefunctions of fermions.
\item Differences also appear in the ionization zones, 
	  and a great deal of them
	  can be attributed to the $\tau$ correction, but not all of it. 
	  The causes for the rest of
	  these differences are not easily identified. 
	  They might be due to the basic
	  differences between the physical- and the chemical approach 
	  to the plasma.
	  The treatment of bound state energies and wave functions
	  might have an effect in this region. 
	  These are highly accurate in MHD but calculated in the isolated
	  particle approximation, whereas they are approximate (fitted to
	  ground-state energies), but varying with the plasma environment
	  in OPAL.
	  We have also
	  tried experimenting with the assumed critical field strength used
	  in MHD for the disruption of a
	  bound state [Eq.\ (4.24) of \citet{mhd1}]. However, this 
	  intervention had only
	  a very small effect.
	  Earlier investigations by \citet{iglesias:OPAL-OP-diff}
	  indicated that a change in the micro-field distribution might have a 
	  greater
	  effect, and that highly excited states are more populated in the 
	  OPAL EOS,
	  although 
	  the OPAL EOS ionizes the plasma more readily than MHD
	  \citep{nayfonov:EOS-high-n}.
\item The evaluation of thermodynamic differentials is done numerically in
	  OPAL but analytically in MHD. For the quantities we have examined here,
	  the difference becomes 
	  most apparent in $\gamma_1$. In the trivial perfect gas region of
	  the $\varrho-T$ plane, OPAL is rugged on a 0.03\%
	  scale (see Sect.\ \ref{diff}), as opposed to the smooth MHD. 
	  These 0.03\% may sound negligible, 
	  but helioseismology is approaching that level.
	  In ionization zones, the discrepancies due to differentiation
	  are most likely larger.
	  On the other hand, physical differences between the two EOS are
	  still at least an order of magnitude larger.
\item New Solar abundance analysis performed on 3D convection simulations by
	  \citet{AGS05} lowers the metallicity and almost halves the oxygen
	  abundance (the most abundant heavy element in the Sun), with detrimental
	  effects
	  on the agreement with helioseismology. Since the helium abundance
	  is found from helioseismic measurements of the bump in $\gamma_1$ from the
	  second helium ionization, and since this feature is blended with the
	  third ionization of oxygen, a reduction in oxygen should be accompanied by
	  an increase in helium. Specifically we find that an 0.45\% increase of $Y$
	  is necessary to keep the bump unchanged with respect to the proposed
	  change of oxygen. This is of the same magnitude, but opposite the change
	  in $Y$ required for a Solar evolution model to fit the current radius and
	  luminosity of the Sun \citep{bahcall:seism-AGS05}.
\end{list}
For helioseismic studies of the equation of state it is a fortunate
property of the Sun that high-$R$ conditions are found exclusively
in the convection zone, where the stratification is essentially adiabatic, 
and therefore virtually decoupled from
radiation and the uncertainty in the
opacity \citep{jcd-wd:osc-eos}. 
As opacity calculations are still subject to errors of 5-10\%,
we stress the importance of the fact
that opacity effects do not contaminate the structure of the convection zone.
This means that the solar convection zone is a perfect
laboratory for investigations of the most controversial parts of the EOS. 

The difference between $\gamma_1$ from and EOS and that of
the Sun can be inferred from helioseismology, and that with an accuracy
that by far exceeds the discrepancy between the two of the best 
present EOS for stellar
structure calculations \citep{jcd-Nature:osc-eos}. The pursuit for a
better EOS is therefore not at all academic, and we can improve both solar
models and atomic physics in the process \citep{sb-jcd:eos-osc}.

\acknowledgements
	We thank J{\o}rgen Christensen--Dalsgaard for supplying us with a copy
	of his solar model S \cite{GONG-Sci:sol-mod}. 
	R.\,T.\ acknowledges support from ARC grant DP\,0342613.
	W.\,D.\ acknowledges support from the grants AST-9987391 and AST-0307578
    of the National Science Foundation.
	In addition, W.D. and R.T were supported in
	part by the Danish National Research Foundation through its
	establishment of the Theoretical Astrophysics Center.


\clearpage

\epsscale{.8}
\clearpage

\clearpage

\clearpage

\clearpage

\clearpage

\clearpage

\clearpage

\clearpage

\clearpage

\clearpage

\clearpage

\clearpage

\clearpage

\clearpage

\clearpage

\clearpage

\clearpage

\clearpage

\clearpage

\clearpage

\clearpage

\clearpage

\clearpage

\clearpage

\epsscale{1}
\clearpage

\clearpage


\end{document}